\documentclass[pra,twocolumn,aps,showpacs,nofootinbib,floatfix,superscriptaddress]{revtex4-1}
\usepackage{amsmath}
\usepackage{amssymb}
\usepackage{graphicx}
\usepackage{color}

\usepackage{braket}

\usepackage{txfonts}

\usepackage[colorlinks=true,linkcolor=blue,citecolor=blue,urlcolor=blue]{hyperref}

\begin{document}

\title{Resolution-enhanced entanglement detection}
%\title{Fisher information and squeezing coefficients for arbitrary-dimensional systems}
%\title{Enhanced entanglement detection through high-resolution imaging}
%\title{Resolution-enhanced detection of multipartite entanglement in arbitrary systems}
\author{Manuel Gessner}
\email{manuel.gessner@ino.it}
\affiliation{QSTAR, INO-CNR and LENS, Largo Enrico Fermi 2, I-50125 Firenze, Italy}
\affiliation{Istituto Nazionale di Ricerca Metrologica, Strada delle Cacce 91, I-10135 Torino, Italy}
\author{Luca Pezz\`e}
\affiliation{QSTAR, INO-CNR and LENS, Largo Enrico Fermi 2, I-50125 Firenze, Italy}
\author{Augusto Smerzi}
\affiliation{QSTAR, INO-CNR and LENS, Largo Enrico Fermi 2, I-50125 Firenze, Italy}
\date{\today}

\begin{abstract}
We formulate a general family of entanglement criteria for multipartite states on arbitrary Hilbert spaces. Fisher information criteria compare the sensitivity to unitary rotations with the variances of suitable local observables. Generalized squeezing-type criteria provide lower bounds that are less stringent but require only measurements of second moments. The enhancement due to local access to the individual subsystems is studied in detail for the case of $N$ spin-$1/2$ particles. The discussed techniques can be readily implemented in current experiments with trapped ions in Paul traps and neutral atoms in optical lattices.
\end{abstract}

\maketitle

\section{Introduction}
Recent progress on atomic and optical experiments has allowed us to design and manipulate many-particle quantum systems at a remarkable level of control. 
A key benchmark for such experiments is the generation and unambiguous detection of quantum entanglement. Entangled quantum states are further required as a key resource for various applications in quantum information and quantum technologies \cite{NielsenChuang,HartmutReview,OBrien2009,LucaRMP,Horodecki2009}.

The difficulty of the entanglement detection problem has spurred the development of a variety of separability criteria \cite{Mintert2005,Horodecki2009,Guhne2009} that do not require full knowledge of the quantum state. Some of these criteria can be formulated in terms of widely accessible observables, rendering them experimentally convenient under fairly general conditions. Prominent examples of entanglement witnesses for spin systems are the Fisher information \cite{PhysRevLett.102.100401,PhysRevA.85.022321,PhysRevA.85.022322,PezzePNAS2016} and the spin-squeezing coefficients \cite{PhysRevA.46.R6797,PhysRevA.47.5138,Sorensen2001,Guhne2009,TothPRA2009,SpinSqueezing,LueckePRL2014}. 
These are successfully used to detect multiparticle entanglement in systems composed of a large number of particles, 
where only global observables are accessible, such as trapped neutral atoms \cite{Esteve2008,Strobel424,AppelPNAS2009,RiedelNATURE2010,LerouxPRL2010a,LuckeSCIENCE2011,BohnetNATPHOT2014,
Peise2015} or ions in Penning traps \cite{Bohnet1297}. %TuraSCIENCE2014,Schmied441

Yet, in many systems, including trapped ions in Paul traps \cite{HartmutReview,Monz,Richerme2014,Jurcevic2014} and cold atoms in optical lattices \cite{Bak2009,She2010,Haller2015,PhysRevLett.114.193001,Islam2015} 
local manipulations and measurements are available and routinely used. 
This raises the question of whether entanglement witnesses based on spin 
squeezing and the Fisher information can be extended to detect a wider class of entangled states by
taking advantage of the local access to the parties~\cite{UshaDeviQIP2003, Gessner2016}.

In this article, we introduce a family of system-independent entanglement criteria, ranging from local to collective addressing. 
Constructing a hierarchy of witness parameters based on the Fisher information and spin squeezing, 
we show how local access can be harnessed to improve the performance of entanglement detection. Some of the entanglement criteria discussed in this paper can further be interpreted in terms of an interferometric quantum advantage for the case of inhomogeneous probing of the subsystems.

This article is organized as follows: In Sec.~\ref{sec:enhanced} we introduce entanglement criteria that make explicit use of local observables~\cite{Gessner2016}. We further discuss different approximations that allow us to extract these criteria from simple measurements. In Sec.~\ref{sec:qubits} the special case of $N$ qubits is studied in detail, and a series of tools for entanglement detection is developed from the general ansatz. 
This includes a locally optimized spin-squeezing coefficient that is especially well suited for entanglement detection. 

\section{Enhanced entanglement detection with local observables}\label{sec:enhanced}
We consider quantum states in an $N$-partite Hilbert space $\mathcal{H}=\mathcal{H}_1\otimes\cdots\otimes\mathcal{H}_N$ with local Hilbert spaces $\mathcal{H}_i$, $i=1,\dots,N$. Separable states are defined as convex combinations of product states \cite{Mintert2005,Horodecki2009,Guhne2009},
\begin{align}\label{eq:fullysep}
\hat{\rho}_{\mathrm{sep}}=\sum_{\gamma}p_{\gamma}\hat{\rho}^{(\gamma)}_1\otimes\cdots\otimes\hat{\rho}^{(\gamma)}_N,
\end{align}
with probabilities $p_{\gamma}$ and quantum states $\hat{\rho}^{(\gamma)}_i$ on the local Hilbert spaces $\mathcal{H}_i$. A necessary condition for separability is given by \cite{Gessner2016}
\begin{align}\label{eq:sepcriterion}
F_Q\bigg[\hat{\rho}_{\mathrm{sep}},\sum_{i=1}^N \hat{A}_i\bigg]&\leq  4\sum_{i=1}^N \mathrm{Var}\big(\hat{A}_i\big)_{\hat{\rho}_{\mathrm{sep}}},
\end{align}
where $\hat{A}_i$ is a local observable acting on the Hilbert space $\mathcal{H}_i$ and $\mathrm{Var}(\hat{H})_{\hat{\rho}}=\langle\hat{H}^2\rangle_{\hat{\rho}} -\langle\hat{H}\rangle_{\hat{\rho}}^2$ denotes the variance of some operator $\hat{H}$ and the quantum expectation values are given as $\langle \hat{H}\rangle_{\hat{\rho}}=\mathrm{Tr}[\hat{H}\hat{\rho}]$. The quantum Fisher information $F_Q[\hat{\rho},\hat{H}]$ quantifies the sensitivity of the initial state $\hat{\rho}$ to unitary transformations $\hat{\rho}(\theta)=e^{-i\hat{H}\theta} \hat{\rho} e^{i\hat{H}\theta}$ \cite{Helstrom1976, paris2009, Giovannetti2011, Varenna, LucaRMP}, which in~(\ref{eq:sepcriterion}) is generated by a sum of local operators $\hat{A}_i$. With the spectral decomposition $\hat{\rho}=\sum_kp_k|\Psi_k\rangle\langle\Psi_k|$, an explicit expression for $F_Q[\hat{\rho},\hat{H}]$ is given by \cite{PhysRevLett.72.3439}
\begin{align}
F_Q[\hat{\rho},\hat{H}]=2\sum_{k,l}\frac{(p_k-p_l)^2}{p_k+p_l}|\langle\Psi_k|\hat{H}|\Psi_l\rangle|^2,
\end{align}
where the sum extends over all pairs with $p_k+p_l\neq 0$. The quantum Fisher information is generally a function of the quantum state and may therefore not always be experimentally accessible without prior information. Nevertheless, it is always possible to obtain the (classical) Fisher information from a recorded probability distribution $p(m|\theta)=\mathrm{Tr}[\hat{\rho}(\theta)\hat{P}_{m}]$, where $\hat{M}$ is an observable with eigenvalues $m$ and corresponding projectors $\hat{P}_{m}$. The Fisher information is then defined as $F_{\hat{M}}[ \hat{\rho}, \hat{H}]=\sum_{m}\frac{1}{p(m|\theta)}\left(\frac{\partial p(m|\theta)}{\partial \theta}\right)^2$ and the quantum Fisher information is the maximum $F_Q[\hat{\rho},\hat{H}]=\max_{\hat{M}}F_{\hat{M}}[ \hat{\rho},\hat{H}]$ \cite{PhysRevLett.72.3439}. Since the inequality~(\ref{eq:sepcriterion}) also holds for the classical Fisher information for arbitrary measurement operators $\hat{M}$, in order to witness entanglement of the state $\hat{\rho}$ it is sufficient to observe $F_{\hat{M}}[\hat{\rho},\sum_{i=1}^N \hat{A}_i]> 4\sum_{i=1}^N \mathrm{Var}\big(\hat{A}_i\big)_{\hat{\rho}}$ for at least one choice of $\hat{M}$ \cite{Gessner2016}. Techniques to extract the Fisher information without full knowledge of the quantum state have been proposed and reported in various different systems \cite{Strobel424,Bohnet1297,PhysRevLett.116.090801,Hauke2016,PezzePNAS2016,Girolami}.

In order to relate the criterion~(\ref{eq:sepcriterion}) to other existing entanglement criteria based on the Fisher information, we note that the local variances on the right-hand side of Eq.~(\ref{eq:sepcriterion}) may be bounded from above by $4 \mathrm{Var}(\hat{A}_i)_{\hat{\rho}}\leq (\lambda_{\max}-\lambda_{\min})^2$, where $\lambda_{\max/\min}$ denote the largest and smallest eigenvalues of $A_i$, respectively. In the case of an $N$-qubit system, the resulting separability bound expresses the so-called shot-noise limit $F_Q[\hat{\rho}_{\mathrm{sep}},\hat{J}_{\mathbf{n}}]\leq  N$, where $\hat{J}_{\mathbf{n}}$ is a collective angular momentum operator \cite{PhysRevLett.102.100401,Varenna}. For unbounded Hilbert-spaces, however, such state-independent bounds no longer exist, as the spectral span is no longer finite.

This limitation is circumvented by the criterion~(\ref{eq:sepcriterion}), which compares the Fisher information with the experimentally measured local variances of the generating operators $\hat{A}_i$ rather than a state-independent upper bound. The additional information provided by these local variances is crucial for two reasons. First, by providing a tight state-dependent upper bound on the Fisher information, it renders the separability criterion more efficient than other criteria that rely on state-independent bounds for these local variances \cite{Gessner2016}. It was shown that for each entangled pure state, a set of local operators $\hat{A}_i$ can be found such that the criterion~(\ref{eq:sepcriterion}) is violated \cite{Gessner2016}. Second, it renders the criterion applicable even in the presence of unbounded Hilbert spaces, where no finite state-independent upper bound exists.

A different, easily accessible entanglement criterion can be derived from Eq.~(\ref{eq:sepcriterion}) by using the upper bound \cite{PhysRevLett.102.100401,Varenna}
\begin{align}\label{eq:FBound}
F_Q[\hat{\rho},\hat{H}_1]\geq \frac{|\langle [\hat{H}_1,\hat{H}_2]\rangle_{\hat{\rho}}|^2}{\mathrm{Var}(\hat{H}_2)_{\hat{\rho}}},
\end{align}
which holds for arbitrary observables $\hat{H}_1$, $\hat{H}_2$ and quantum states $\hat{\rho}$ on $\mathcal{H}$. Separable states $\hat{\rho}_{\mathrm{sep}}$ must necessarily satisfy the condition \cite{Gessner2016}
\begin{align}\label{eq:variancecriterion}
\mathrm{Var}(\hat{A})_{\Pi(\hat{\rho}_{\mathrm{sep}})}\mathrm{Var}(\hat{B})_{\hat{\rho}_{\mathrm{sep}}}\geq \frac{|\langle [\hat{A},\hat{B}]\rangle_{\hat{\rho}_{\mathrm{sep}}}|^2}{4},
\end{align}
where $\hat{A}=\sum_{i=1}^N\hat{A}_i$ is a sum of local operators and $\hat{B}$ is an arbitrary operator. 
Moreover, we have rewritten the sum over the local variances $\sum_{i=1}^N\mathrm{Var}\big(\hat{A}_i\big)_{\hat{\rho}}=\mathrm{Var}(\sum_{i=1}^N\hat{A}_i)_{\Pi(\hat{\rho})}$ by constructing a product state $\Pi(\hat{\rho})=\hat{\rho}_1\otimes\cdots\otimes\hat{\rho}_N$ from the reduced density operators $\hat{\rho}_i$ of $\hat{\rho}$ \cite{Gessner2016}. The separability criterion~(\ref{eq:variancecriterion}) is accessible by measurements of moments up to second order. Note that despite its formal resemblance, the derivation of the above criterion does not involve any uncertainty relation. In fact, Heisenberg's uncertainty relation follows from Eq.~(\ref{eq:FBound}) since $F_Q[\hat{\rho},\hat{H}]\leq 4\mathrm{Var}(\hat{H})_{\hat{\rho}}$ always holds \cite{PhysRevLett.72.3439}. 

The separability criteria~(\ref{eq:sepcriterion}) and (\ref{eq:variancecriterion}) suggest the introduction of two general classes of coefficients. Specifically, we introduce the \textit{generalized variance-assisted Fisher densities}
\begin{align}\label{eq:genFD}
f(\hat{\rho})=\frac{F_Q[\hat{\rho},\hat{A}]}{4\mathrm{Var}(\hat{A})_{\Pi(\hat{\rho})}}
\end{align}
and \textit{squeezing coefficients}
\begin{align}\label{eq:genSC}
\xi^2(\hat{\rho})=\frac{4\mathrm{Var}(\hat{A})_{\Pi(\hat{\rho})}\mathrm{Var}(\hat{B})_{\hat{\rho}}}{|\langle [\hat{A},\hat{B}]\rangle_{\hat{\rho}}|^2}
\end{align}
for \textit{arbitrary} multipartite systems. Recall that $\hat{A}=\sum_{i=1}^N\hat{A}_i$ and $\hat{B}$ is arbitrary. According to Eqs.~(\ref{eq:sepcriterion}) and~(\ref{eq:variancecriterion}), any observation of $f(\hat{\rho})> 1$ and $\xi^{-2}(\hat{\rho})> 1$ indicates entanglement of the state $\rho$. Notice that by virtue of Eq.~(\ref{eq:FBound}), we have
\begin{align}
f(\hat{\rho})\geq \xi^{-2}(\hat{\rho}).
\end{align}
All these bounds hold for arbitrary choices of the local operators $\hat{A}_i$, as well as $\hat{B}$. Hence, these operators can be adjusted in order to maximize the quantities $f(\hat{\rho})$ and $\xi^{-2}(\hat{\rho})$ for a given $\hat{\rho}$. Such suitably optimized coefficients are employed as a central tool for entanglement detection in this article. 

\section{$N$-qubit systems}\label{sec:qubits}
In this section, we discuss applications of the coefficients~(\ref{eq:genFD}) and (\ref{eq:genSC}) to systems of $N$ qubits, i.e., $\mathcal{H}_1=\cdots=\mathcal{H}_N=\mathbb{C}^2$, based on local or global optimizations of the operators $\hat{A}$ and $\hat{B}$, under different constraints.

\subsection{Local optimizations of $N$-qubit systems: General considerations}\label{sec:generallocalqubit}
In an $N$-qubit system, the local operators $\hat{A}_i$, which appear in the argument of the Fisher information on the left-hand side of Eq.~(\ref{eq:sepcriterion}), can be parametrized in terms of Pauli matrices $\hat{\boldsymbol{\sigma}}_i=(\hat{\sigma}_i^{(x)},\hat{\sigma}^{(y)}_i,\hat{\sigma}^{(z)}_i)$ as
\begin{align}\label{eq:AofC}
\hat{A}(\mathbf{c})=\frac{1}{2}\sum_{i=1}^N\mathbf{n}_i\cdot\hat{\boldsymbol{\sigma}}_i,
\end{align}
where $\mathbf{c}=(\mathbf{n}_1,\dots,\mathbf{n}_N)$ and $\mathbf{n}_i\in\mathbb{R}^3$ are local vectors. %\footnote{For ease of notation we restrain the usage of transposition symbols in the definition of vectors. Henceforth definitions of composed vectors are interpreted as column vectors with elements $\mathbf{c}=(\mathbf{n}_1,\dots,\mathbf{n}_N)=(n^{(x)}_1,n^{(y)}_1,n^{(z)}_1,\dots,n^{(z)}_N,n^{(y)}_N,n^{(z)}_N)$.} 
The vectors $\mathbf{n}_i$ may not only give different orientation to the local spins but they may also attribute different weights.
Equation~(\ref{eq:AofC}) thus formally corresponds to a weighted spin operator.
%Note that $|\mathbf{c}|^2=N$ when all $\mathbf{n}_i$ are unit vectors. 
To optimize the vectors $\mathbf{n}_i$ locally, such that the quantum Fisher information attains its maximum value, we employ the $3N\times 3N$ local quantum Fisher matrix, defined elementwise as
\begin{align}
\left(\mathbf{Q}^{L}_{\hat{\rho}}\right)^{\alpha\beta}_{ij}=\frac{1}{2}\sum_{k,l}\frac{(p_k-p_l)^2}{p_k+p_l}\langle\Psi_k|\hat{\sigma}_i^{(\alpha)}|\Psi_l\rangle\langle\Psi_l|\hat{\sigma}_j^{(\beta)}|\Psi_k\rangle,
\end{align}
on the basis of the spectral decomposition $\hat{\rho}=\sum_kp_k|\Psi_k\rangle\langle\Psi_k|$. The indices $i,j\in\{1,\dots,N\}$ refer to different subsystems and $\alpha,\beta\in\{x,y,z\}$ indicate the orientation. Using this matrix, we can write \cite{PhysRevA.82.012337}
\begin{align}
F_Q[\hat{\rho},\hat{A}(\mathbf{c})]=\mathbf{c}^T\mathbf{Q}^{L}_{\hat{\rho}}\mathbf{c},
\end{align}
showing that the maximum value of the quantum Fisher information is given by the largest eigenvalue of $\mathbf{Q}^{L}_{\hat{\rho}}$ and the optimal operator $A(\mathbf{c})$ is given by Eq.~(\ref{eq:AofC}) when $\mathbf{c}$ is the eigenvector corresponding to the largest eigenvalue. Using the local covariance matrix 
\begin{align}
(\boldsymbol{\Gamma}^{L}_{\hat{\rho}})^{\alpha\beta}_{ij}=\frac{1}{4}\mathrm{Cov}(\hat{\sigma}_i^{(\alpha)},\hat{\sigma}_j^{(\beta)})_{\hat{\rho}}
\end{align}
where $\mathrm{Cov}(\hat{A},\hat{B})_{\hat{\rho}}=\langle\frac{1}{2}\{\hat{A},\hat{B}\}\rangle_{\hat{\rho}}-\langle A\rangle_{\hat{\rho}}\langle B\rangle_{\hat{\rho}}$  with the anticommutator $\{\hat{A},\hat{B}\}=\hat{A}\hat{B}+\hat{B}\hat{A}$, we arrive at the analogous expression
\begin{align}
\mathrm{Var}(\hat{A}(\mathbf{c}))_{\hat{\rho}}=\mathbf{c}^{T}\boldsymbol{\Gamma}^{L}_{\hat{\rho}}\mathbf{c}.
\end{align}
Hence, the strongest separability criterion based on Eq.~(\ref{eq:sepcriterion}) is given by the largest eigenvalue of the matrix $\mathbf{Q}^{L}_{\hat{\rho}}-4\boldsymbol{\Gamma}^{L}_{\Pi(\hat{\rho})}$, which is necessarily smaller than or equal to zero for all separable states \cite{Gessner2016}. This condition is, in fact, \textit{necessary and sufficient} for separability of pure $N$-qubit states \cite{Gessner2016}.

Next, we derive a bound for the sum of local variances that appears on the right-hand side of Eq.~(\ref{eq:sepcriterion}). Each variance in the sum is determined by the respective reduced density matrix $\hat{\rho}_i$ of the state $\hat{\rho}$, where $i=1,\dots,N$. The single-qubit space is conveniently described by the Bloch-vector picture. Introducing
\begin{align}\label{eq:localBlochVectors}
\hat{\rho}_i=\frac{1}{2}(\hat{\mathbb{I}}_2+\mathbf{m}_i\cdot\hat{\boldsymbol{\sigma}}_i),
\end{align}
where $\mathbf{m}_i$ denotes the local Bloch vector and $\hat{\mathbb{I}}_2$ is the identity operator on $\mathbb{C}^2$, the local variances are given by
\begin{align}\label{eq:localvariancesBloch}
4\mathrm{Var}\big(\hat{A}(\mathbf{c})\big)_{\Pi(\hat{\rho})}=\sum_{i=1}^N\mathrm{Var}\big(\mathbf{n}_i\cdot\hat{\boldsymbol{\sigma}}_i\big)_{\hat{\rho}} = |\mathbf{c}|^2 - \sum_{i=1}^N(\mathbf{n}_i\cdot\mathbf{m}_i)^2,
\end{align}
where $|\mathbf{c}|^2=\sum_{i=1}^N|\mathbf{n}_i|^2$. The quantity~(\ref{eq:localvariancesBloch}) is minimized by choosing $\mathbf{n}_i\parallel\mathbf{m}_i$ for all $i$. Note that the length of the Bloch vector is related to the states' purity, $|\mathbf{m}_i|^2=\mathrm{Tr}\{\hat{\rho}_i^2\}\leq 1$, where equality is reached if and only if the local state is pure, i.e., $\hat{\rho}_i=|\Psi_i\rangle\langle\Psi_i|$. Hence, only when all reduced states $\hat{\rho}_i$ for $i=1,\dots,N$ are pure (which implies that the state of the $N$-particle system is an uncorrelated product state $|\Psi_1\rangle\otimes\dots\otimes|\Psi_N\rangle$) is it possible to choose an operator $\hat{A}(\mathbf{c})$ for which the local variances vanish. For this particular choice the quantum Fisher information $F_{Q}[|\Psi_1\rangle\otimes\cdots\otimes|\Psi_N\rangle,\sum_{i=1}^N\mathbf{m}_i\cdot\hat{\boldsymbol{\sigma}}_i/2]$ also vanishes. In all other cases the sum over local variances yields a finite value. Finally, we note that the upper bound
\begin{align}\label{eq:upperboundLclVar}
4\mathrm{Var}\big(\hat{A}(\mathbf{c})\big)_{\Pi(\hat{\rho})}\leq |\mathbf{c}|^2
\end{align}
can always be saturated for arbitrary states: when $|\mathbf{m}_i|>0$, the variance is maximized by choosing $\mathbf{n}_i\perp\mathbf{m}_i$, whereas if $|\mathbf{m}_i|=0$, the subsystem is maximally mixed, and the variance is maximal for arbitrary local orientations. %Note from Eq.~(\ref{eq:localvariancesBloch}) that the bound~(\ref{eq:upperboundLclVar}) is always valid, even if the $\mathbf{n}_i$ are not unit vectors.

\subsection{Locally optimized quantum Fisher densities}\label{sec:lclFD}
Using Eq.~(\ref{eq:AofC}), we can rewrite the separability criterion~(\ref{eq:sepcriterion}) as $f^V_{\mathbf{c}}(\hat{\rho})\leq 1$, where
\begin{align}\label{eq:fcv}
f^V_{\mathbf{c}}(\hat{\rho}):=\frac{F_Q[\hat{\rho},\hat{A}(\mathbf{c})]}{4\mathrm{Var}\big(\hat{A}(\mathbf{c})\big)_{\Pi(\hat{\rho})}}
\end{align}
defines a variance-assisted quantum Fisher density [recall Eq.~(\ref{eq:genFD})] \cite{footnote1}. In the remainder of this article, those coefficients which require knowledge of the local variances carry the superscript $V$. The condition $f^V_{\mathbf{c}}(\hat{\rho})>1$, for any choice of $\mathbf{c}$, indicates entanglement of $\hat{\rho}$. Note that the coefficient $f^V_{\mathbf{c}}(\hat{\rho})$ does not depend on the total length of the vector $\mathbf{c}$ and only on the relative strengths and directions of its local components.

We maximize Eq.~(\ref{eq:fcv}) by taking the maximal eigenvector $\mathbf{c}^L_{\max}$ of the matrix $\mathbf{Q}^{L}_{\hat{\rho}}-4\boldsymbol{\Gamma}^{L}_{\Pi(\hat{\rho})}$ introduced before \cite{footnote2}. This leads to the most powerful separability condition discussed in this section, i.e., $f^V_L(\hat{\rho})\leq 1$, where
\begin{align}\label{eq:flv}
f^V_L(\hat{\rho}):=\frac{F_Q[\hat{\rho},\hat{A}(\mathbf{c}^{L}_{\max})]}{4\mathrm{Var}\left(\hat{A}(\mathbf{c}^{L}_{\max})\right)_{\Pi(\hat{\rho})}},
\end{align}
$\mathrm{Var}(\hat{A}(\mathbf{c}^{L}_{\max}))_{\Pi(\hat{\rho})}=(\mathbf{c}^{L}_{\max})^T\boldsymbol{\Gamma}^{L}_{\Pi(\hat{\rho})}\mathbf{c}^{L}_{\max}=\sum_{i=1}^N\mathrm{Var}(\mathbf{n}^{L}_{\max,i}\cdot\hat{\boldsymbol{\sigma}}_{i})_{\hat{\rho}},$ and $\mathbf{c}^{L}_{\max}=(\mathbf{n}^{L}_{\max,1},\dots,\mathbf{n}^{L}_{\max,N})$. The results of Ref.~\cite{Gessner2016} imply that the condition $f^V_L(|\Psi\rangle)>1$ is necessary and sufficient for entanglement of the pure state $|\Psi\rangle$.

A simpler coefficient, albeit one that produces a less stringent separability criterion, is obtained when replacing the sum of local variances by its the upper bound~(\ref{eq:upperboundLclVar}). In this case it is more suitable to maximize the value of the quantum Fisher information by identifying the eigenvector $\mathbf{c}^{L}_{\mathrm{opt}}$ which leads to the maximum eigenvalue $\lambda^{L}_{\mathrm{opt}}$ of the matrix $\mathbf{Q}^{L}_{\hat{\rho}}$. We obtain the Fisher density
\begin{align}\label{eq:fln}
f_L(\hat{\rho}):=\frac{F_Q[\hat{\rho},\hat{A}(\mathbf{c}^{L}_{\mathrm{opt}})]}{|\mathbf{c}^L_{\mathrm{opt}}|^2}\equiv \lambda^L_{\mathrm{opt}}.
\end{align}
where $F_Q[\hat{\rho},\hat{A}(\mathbf{c}^{L}_{\mathrm{opt}})]=|\mathbf{c}^{L}_{\mathrm{opt}}|^2\lambda^{L}_{\mathrm{opt}}$ and $|\mathbf{c}^{L}_{\mathrm{opt}}|^2=\sum_{i=1}^N|\mathbf{n}_{\mathrm{opt},i}^L|^2$.

Note that, in general, the obtained collective vectors $\mathbf{c}^{L}_{\max}$ and $\mathbf{c}^{L}_{\mathrm{opt}}$ do not lead to locally normalized vectors $\mathbf{n}_i$ in Eq.~(\ref{eq:AofC}); that is, the unconstrained optimization may assign more weight to certain qubits with respect to others, and we may find 
$|\mathbf{n}^{L}_{\max,i}|^2\neq 1$ and $|\mathbf{n}^{L}_{\mathrm{opt},i}|^2\neq 1$. This means that the optimal local operator $\hat{A}(\mathbf{c})$ may be attained when the couplings to the local subsystems are of unequal strengths; see also Sec.~\ref{sec:metro} for a discussion in the context of high-precision phase estimation.

This justifies the introduction of an additional coefficient,
\begin{align}\label{eq:flnn}
f_l(\hat{\rho}):=\max_{\substack{\{\mathbf{n}_{1},\dots,\mathbf{n}_{N}\}\\|\mathbf{n}_i|^2=1}}\frac{F_Q[\hat{\rho},\frac{1}{2}\sum_{i=1}^N\mathbf{n}_i\cdot\hat{\boldsymbol{\sigma}}_i]}{N},
\end{align}
where the quantum Fisher information is optimized with equal weight on each qubit: $|\mathbf{n}_i|^2=1$ for all $i=1,\dots,N$. This implies $\sum_{i=1}^N|\mathbf{n}_{i}|^2=N$. These $N$ constraints reduce the optimization to an effectively $2N$-dimensional parameter space. Henceforth, the subscript $l$ will be used to label quantities that are obtained from such a constrained local optimization. An individual optimization of the local qubit generators to maximize the Fisher information was discussed already in Ref.~\cite{PhysRevA.82.012337}. This coefficient quantifies the metrological quantum gain of the input state $\hat{\rho}$ and is able to assess the number of parties that are entangled with each other \cite{PhysRevA.85.022321}; further details and comparison to other coefficients will be provided in Secs.~\ref{sec:generalproperties}, \ref{sec:metro}, and \ref{sec:multi}.

\subsection{Globally optimized quantum Fisher densities}
Let us now consider the case where all local directions are chosen to be equal and normalized, i.e., $\mathbf{n}_1=\cdots=\mathbf{n}_N=\mathbf{n}$, with $|\mathbf{n}|^2=1$. The operator $\hat{A}(\mathbf{c})$, Eq.~(\ref{eq:AofC}), now reduces to
\begin{align}
\hat{J}_{\mathbf{n}}=\mathbf{n}\cdot\hat{\mathbf{J}},
\end{align}
where $\hat{\mathbf{J}}=(\hat{J}_x,\hat{J}_y,\hat{J}_z)$ is composed of collective angular momentum operators $\hat{J}_{\alpha}=\sum_{i=1}^N\hat{\sigma}^{(\alpha)}_i/2$, with $\alpha=x,y,z$. Following a procedure analogous to the one we used before, we now obtain the separability condition $f^V_G(\hat{\rho})\leq 1$, where the globally maximized, variance-assisted Fisher density is given by
\begin{align}\label{eq:fgv}
f^V_G(\hat{\rho}):=\frac{F_Q[\hat{\rho},\hat{J}_{\mathbf{n}_{\max}}]}{4\mathrm{Var}(\hat{J}_{\mathbf{n}_{\max}})_{\Pi(\hat{\rho})}}
\end{align}
and $4\mathrm{Var}(\hat{J}_{\mathbf{n}_{\max}})_{\Pi(\hat{\rho})}=\sum_{i=1}^N\mathrm{Var}(\mathbf{n}_{\max}\cdot\hat{\boldsymbol{\sigma}}_i)_{\hat{\rho}}$. The unit vector $\mathbf{n}_{\max}\in\mathbb{R}^3$ yields the maximum eigenvalue of the $3\times 3$ matrix $\mathbf{Q}^{G}_{\hat{\rho}}-4\boldsymbol{\Gamma}^{G}_{\hat{\rho}}$, where the global quantum Fisher matrix
\begin{align}\label{eq:globalQFM}
\left(\mathbf{Q}^{G}_{\hat{\rho}}\right)_{\alpha\beta}=2\sum_{k,l}\frac{(p_k-p_l)^2}{p_k+p_l}\langle\Psi_k|\hat{J}_{\alpha}|\Psi_l\rangle\langle\Psi_l|\hat{J}_{\beta}|\Psi_k\rangle
\end{align}
and the global covariance matrix
\begin{align}
(\boldsymbol{\Gamma}^{G}_{\hat{\rho}})_{\alpha\beta}=\mathrm{Cov}(\hat{J}_{\alpha},\hat{J}_{\beta})_{\hat{\rho}}
\end{align}
replace the $3N\times 3N$ matrices that were used for the local optimization in Sec.~\ref{sec:lclFD}. We introduced the label $G$ to distinguish global quantities in $\mathbb{R}^3$ from those with local resolution in $\mathbb{R}^{3N}$, which were labeled by $L$.

Finally, in analogy to Eq.~(\ref{eq:fln}), we can replace the variance in the denominator 
of Eq.~(\ref{eq:fgv}) by its upper bound~(\ref{eq:upperboundLclVar}). Here, the vector $\mathbf{c}\in\mathbb{R}^{3N}$ in Eq.~(\ref{eq:upperboundLclVar}) contains $N$ identical copies of a unit vector $\mathbf{n}\in\mathbb{R}^{3}$ and consequently leads to $|\mathbf{c}|^2=N$. The maximum 
eigenvalue $\lambda^{G}_{\mathrm{opt}}$ and associated eigenvector 
$\mathbf{n}_{\mathrm{opt}}$ of $\mathbf{Q}^{G}_{\hat{\rho}}$ yield the Fisher density under global optimizations \cite{PhysRevLett.102.100401, Varenna}
\begin{align}\label{eq:fgn}
f_G(\hat{\rho}):=\frac{F_Q[\hat{\rho},\hat{J}_{\mathbf{n}_{\mathrm{opt}}}]}{N},
\end{align}
where $F_Q[\hat{\rho},\hat{J}_{\mathbf{n}_{\mathrm{opt}}}]=\lambda^{G}_{\mathrm{opt}}$. 

\subsection{Locally optimized spin-squeezing coefficients}\label{sec:spinsqueezing}
We now turn to separability bounds based on Eq.~(\ref{eq:variancecriterion}):
\begin{align}\label{eq:squeezingbound}
\xi^{-2}_{\mathbf{c},\hat{B}}(\hat{\rho})\leq 1
\end{align}
holds for all separable states, where
\begin{align}\label{eq:localvariancecoefficient}
\xi^2_{\mathbf{c},\hat{B}}(\hat{\rho}):=\frac{4\mathrm{Var}(\hat{A}(\mathbf{c}))_{\Pi(\hat{\rho})}\mathrm{Var}(\hat{B})_{\hat{\rho}}}{|\langle [\hat{A}(\mathbf{c}),\hat{B}]\rangle_{\hat{\rho}}|^2}.
\end{align}
Note that the operator $\hat{B}$ is arbitrary, while $\hat{A}(\mathbf{c})$ was defined in Eq.~(\ref{eq:AofC}). We can optimize the coefficient $\xi_{\mathbf{c},\hat{B}}$ conveniently if we restrict ourselves to linear operators $\hat{B}=\hat{A}(\mathbf{c}_2)=\sum_{i=1}^N\mathbf{n}'_i\cdot\hat{\boldsymbol{\sigma}}_i/2$, with $\mathbf{c}_2=(\mathbf{n}'_1,\dots,\mathbf{n}'_N)$. We first notice that for an arbitrary pair $\mathbf{c}_1=(\mathbf{n}_1,\dots,\mathbf{n}_N)$ and $\mathbf{c}_2$ the following relation holds:
\begin{align}\label{eq:AofCCommutators}
[\hat{A}(\mathbf{c}_1),\hat{A}(\mathbf{c}_2)]%&=\frac{1}{4}\sum_{i,j=1}^N\sum_{\alpha,\beta=x,y,z}n^{(\alpha)}_in^{\prime(\beta)}_j[\sigma_i^{(\alpha)},\sigma_j^{(\beta)}]\\
&=\frac{i}{2}\sum_{i=1}^N(\mathbf{n}_i\times\mathbf{n}'_i)\cdot\boldsymbol{\sigma}_i%\\&
=i\hat{A}(\mathbf{c}_3),
\end{align}
where $\mathbf{c}_3=(\mathbf{n}_i\times\mathbf{n}'_1,\dots,\mathbf{n}_N\times\mathbf{n}'_N)$. The coefficient~(\ref{eq:localvariancecoefficient}) can thus be expressed as
\begin{align}\label{eq:localspinsqueezevectors}
\xi^2_{\mathbf{c}_1,\mathbf{c}_2}(\hat{\rho}):=\frac{4\mathrm{Var}(\hat{A}(\mathbf{c}_1))_{\Pi(\hat{\rho})}\mathrm{Var}(\hat{A}(\mathbf{c}_2))_{\hat{\rho}}}{|\langle \hat{A}(\mathbf{c}_3)\rangle_{\hat{\rho}}|^2},
\end{align}
where the vectors $\mathbf{c}_1$ and $\mathbf{c}_2$ determine $\mathbf{c}_3$. Special cases of Eq.~(\ref{eq:localspinsqueezevectors}) are obtained by optimizing the vectors $\mathbf{c}_1$ and $\mathbf{c}_2$ with different constraints. In the following, we present a local spin-squeezing coefficient based on a locally normalized approach; that is, all of the qubits are considered with equal weight. Two possible inhomogeneous generalizations are discussed in Appendix~\ref{app:lclss}.

It is convenient to introduce the following definitions: We call any vectors $\mathbf{c}_1$, $\mathbf{c}_2$ \textit{locally orthogonal} if their local components satisfy $\mathbf{n}_i\perp\mathbf{n}'_i$ for all $i=1,\dots,N$. Moreover, vectors $\mathbf{c}_1$ that satisfy $|\mathbf{n}_i|^2=1$ for all $i=1,\dots,N$ are called \textit{locally normalized}. A local spin-squeezing coefficient can be obtained from Eq.~(\ref{eq:localspinsqueezevectors}) by choosing a pair of locally orthogonal and locally normalized vectors $\mathbf{c}_1$ and $\mathbf{c}_2$, such that $\mathbf{c}_3$ coincides with the local mean-spin vector
\begin{align}\label{eq:localmeanspin}
\mathbf{c}_0%&=\left(\frac{\langle \hat{\boldsymbol{\sigma}}_1\rangle_{\hat{\rho}}}{\left|\langle \hat{\boldsymbol{\sigma}}_1\rangle_{\hat{\rho}}\right|},\dots,\frac{\langle \hat{\boldsymbol{\sigma}}_N\rangle_{\hat{\rho}}}{\left|\langle \hat{\boldsymbol{\sigma}}_N\rangle_{\hat{\rho}}\right|}\right)\\
&=\left(\frac{ \mathbf{m}_1}{\left|\mathbf{m}_1\right|},\dots,\frac{\mathbf{m}_N}{\left|\mathbf{m}_N\right|}\right),
\end{align}
where, according to Eq.~(\ref{eq:localBlochVectors}), $\mathbf{m}_i=\langle\hat{\boldsymbol{\sigma}}_i\rangle_{\hat{\rho}}$ is a vector with components given by the local spin expectation values. According to Eq.~(\ref{eq:AofCCommutators}), the condition $\mathbf{c}_3=\mathbf{c}_0$ requires that both $\mathbf{c}_1$ and $\mathbf{c}_2$ are locally orthogonal to $\mathbf{c}_0$, i.e., $\mathbf{n}_i,\mathbf{n}'_i\perp\mathbf{m}_i$ for all $i=1,\dots,N$. From Eq.~(\ref{eq:localvariancesBloch}), we obtain
\begin{align}\label{eq:localVarOrthSpinSqueeze}
4\mathrm{Var}\big(\hat{A}(\mathbf{c}_1)\big)_{\Pi(\hat{\rho})}= |\mathbf{c}_1|^2 - \sum_{i=1}^N(\mathbf{n}_i\cdot\mathbf{m}_i)^2 =|\mathbf{c}_1|^2.
\end{align}
Consequently, replacing the local variances by their upper bound~(\ref{eq:upperboundLclVar}) does not imply a loss of generality for any vector $\mathbf{c}_1$ that is locally orthogonal to the mean spin $\mathbf{c}_0$. Together with the local normalization condition $|\mathbf{n}_i|^2=1$ we now obtain $4\mathrm{Var}\big(\hat{A}(\mathbf{c}_1)\big)_{\Pi(\hat{\rho})}=N$. This leads to the local spin-squeezing coefficient
\begin{align}\label{eq:xil}
\xi_l^2(\hat{\rho}):=\min_{\mathbf{c}_{\perp}}\frac{N\mathrm{Var}(\hat{A}(\mathbf{c}_{\perp}))_{\hat{\rho}}}{\langle \hat{A}(\mathbf{c}_0)\rangle_{\hat{\rho}}^2},
\end{align}
where the optimization is constrained to vectors $\mathbf{c}_{\perp}$ which are locally normalized and locally orthogonal to the mean-spin vector $\mathbf{c}_0$. Effectively, these constraints reduce the number of free parameters in the optimization from $3N$ to $N$. Here, the spin expectation value reads
\begin{align}\label{eq:normmeanspinexpval}
\langle\hat{A}(\mathbf{c}_0)\rangle_{\hat{\rho}}=\frac{1}{2}\sum_{i=1}^N\frac{\mathbf{m}_i}{|\mathbf{m}_i|}\cdot\langle\hat{\boldsymbol{\sigma}}_i\rangle_{\hat{\rho}}=\frac{1}{2}\sum_{i=1}^N|\mathbf{m}_i|.
\end{align}
The coefficient $\xi_l$ coincides with the one introduced in Ref.~\cite{UshaDeviQIP2003}, where it was derived as a direct generalization of Ref.~\cite{Sorensen2001}. We remark that the coefficient $\xi_l$ is only well defined when none of the local spins is maximally mixed, i.e., $|\mathbf{m}_i|\neq 0$; otherwise, the locally normalized mean-spin vector~(\ref{eq:localmeanspin}) is undefined. As is shown in Appendix~\ref{app:lclss}, the inhomogeneous coefficients $\xi_L$ and $\xi_{Ll}$, defined in Eqs.~(\ref{eq:xiL}) and~(\ref{eq:xiLl}), respectively, do not suffer from this limitation and represent stronger entanglement criteria than $\xi_l$.

\subsection{Globally optimized spin-squeezing coefficients}
Considering only collective orientations $\mathbf{n}\in\mathbb{R}^3$, i.e., choosing $\mathbf{n}_1=\cdots=\mathbf{n}_N\equiv\mathbf{n}$, with $|\mathbf{n}_i|^2=1$, we obtain the spin-squeezing coefficients
\begin{align}\label{eq:xign}
\xi_G^2(\hat{\rho}):=\min_{\mathbf{n}_{\perp}}\frac{N\mathrm{Var}(\hat{J}_{\mathbf{n}_{\perp}})_{\hat{\rho}}}{\langle \hat{J}_{\mathbf{n}_0}\rangle_{\hat{\rho}}^2}
\end{align}
and
\begin{align}\label{eq:xigv}
\left(\xi^V_G\right)^2(\hat{\rho}):=\min_{\mathbf{n}_{\perp}}\frac{4\mathrm{Var}(\hat{J}_{\mathbf{n}'_{\perp}})_{\Pi(\hat{\rho})}\mathrm{Var}(\hat{J}_{\mathbf{n}_{\perp}})_{\hat{\rho}}}{\langle \hat{J}_{\mathbf{n}_0}\rangle_{\hat{\rho}}^2},
\end{align}
depending on whether we use the local variance or their upper bound~(\ref{eq:upperboundLclVar}), respectively, which generally may yield different results. Here,
\begin{align}\label{eq:globalmeanspin}
\mathbf{n}_0 = \langle \hat{\mathbf{J}}\rangle_{\hat{\rho}}/|\langle \hat{\mathbf{J}}\rangle_{\hat{\rho}}|
\end{align}
defines the global mean-spin direction and $\{\mathbf{n}_0,\mathbf{n}_{\perp},\mathbf{n}'_{\perp}\}$ are mutually orthogonal directions. When the state is symmetric under the exchange of subsystems, we find $\mathbf{n}_0=\mathbf{m}_i/|\mathbf{m}_i|$ for $i=1,\dots,N$. 

The parameter $\xi_G$ corresponds to the spin-squeezing coefficient introduced by Wineland \textit{et al.} to quantify the squeezing-enabled enhancement of the phase sensitivity for Ramsey interferometry \cite{PhysRevA.46.R6797,PhysRevA.50.67}. All of the coefficients introduced in this section are special cases of $\xi^2_{\mathbf{c},\hat{B}}$; hence, for arbitrary separable states, they must satisfy the upper bound~(\ref{eq:squeezingbound}).

A powerful extension of the coefficients proposed here can be achieved by considering nonlinear operators $\hat{B}$ in Eq.~(\ref{eq:localvariancecoefficient}) \cite{LuckeSCIENCE2011,PhysRevA.92.012102}. In Ref.~\cite{PhysRevA.92.012102} saturation of inequality~(\ref{eq:FBound}) was demonstrated for specific states and optimally chosen pairs of operators $\hat{A}$ and $\hat{B}$. However, these operators $\hat{A}$ do not necessarily yield the maximum Fisher information $F_Q[\hat{\rho},\hat{A}]$ for a given state $\hat{\rho}$.

%The spin-squeezing coefficients, such as $\xi_G$ are widely employed to detect entangled quantum states in experiments \cite{Esteve2008,LucaRMP}, and theoretical studies \cite{SpinSqueezing}. The complete set of separability criteria which can be obtained from measurements of collective spin operators has been characterized in Ref.~\cite{TothPRA2009}. %Our goal is here to study when additional access to the expectation values and variances of local spin observables can enhance the entanglement detection efficiency.

\subsection{Comparison of local and global coefficients:\\ General considerations}\label{sec:generalproperties}
\begin{table*}[t]
\begin{tabular}{l|cc|c|c|c|cccc}
 \quad & \quad Defining \quad & \quad Fisher \quad & \multicolumn{3}{|c|}{Optimization} & \quad Local variances \quad & \quad $k$ \quad & \quad Ref. \\ 
  &  \quad Eq.\quad  & information \quad & Space & Constraints & Free parameters \quad &  & separability &  \\ \hline
$f^V_L$ & (\ref{eq:flv}) & Yes & $\mathbb{R}^{3N}$ & $1$ & $3N-1$ & Yes & No & \\
$f^V_G$ & (\ref{eq:fgv}) & Yes & $\mathbb{R}^{3}$ & $1$ & $2$ & Yes & No &  \\
$f_L$ & (\ref{eq:fln}) & Yes & $\mathbb{R}^{3N}$ & $1$ & $3N-1$ & No & No & \\
$f_l$ & (\ref{eq:flnn}) & Yes & $\mathbb{R}^{3N}$ & $N$ & $2N$ & No & Yes & \cite{PhysRevA.82.012337} \\
$f_G$ & (\ref{eq:fgn}) & Yes & $\mathbb{R}^{3}$ & $1$ & $2$ & No & Yes & \cite{PhysRevLett.102.100401,PhysRevA.82.012337} \\
%$\xi_L$ & (\ref{eq:xiL}) & No & $\mathbb{R}^{3N}$ & $N$ & $2N$ & No (always max.) & No & \\
%$\xi_{Ll}$ & (\ref{eq:xiLl}) & No & $\mathbb{R}^{3N}$ & $2N$ & $N$ & No (always max.) & No & \\
$\xi_l$ & (\ref{eq:xil}) & No & $\mathbb{R}^{3N}$ & $2N$ & $N$ & No (always maximal) & Yes & \cite{UshaDeviQIP2003} \\
$\xi^V_G$ & (\ref{eq:xigv}) & No & $\mathbb{R}^{3}$ & $1$ & $2$ & Yes & No &  \\
$\xi_G$ & (\ref{eq:xign}) & No & $\mathbb{R}^{3}$ & $1$ & $2$ & No & Yes & \cite{PhysRevA.46.R6797,PhysRevA.50.67,Sorensen2001} \\
\end{tabular}
\caption{Comparison of the entanglement coefficients introduced in Sec.~\ref{sec:qubits}. The Fisher densities $f$ compare the Fisher information to suitable bounds; the spin-squeezing coefficients $\xi$ require only measurements of variances and mean values. The space, constraints, and free parameters involved in the optimization are summarized in the middle part of the table. Coefficients with subscripts $L$ and $l$ are obtained by maximizing the witness over all $N$ local directions, leading to a $3N$-dimensional real space, subject to different constraints. Those with subscript $G$ are obtained by optimizing only over one global spin direction in $\mathbb{R}^3$. All optimizations except for those of $f_l$ and $\xi_l$ correspond to finding the largest eigenvalue and eigenvector of a matrix. Coefficients with superscript $V$ make explicit use of the value of the local variances, whereas those without a superscript compare to an upper bound. In the case of $\xi_l$ it was proven that the local variances always saturate their upper bound; for generalizations of $\xi_l$ we refer to Appendix~\ref{app:lclss}. As explained in Sec.~\ref{sec:multi}, certain coefficients are able to provide bounds on the $k$-separability class of the state. Some of these coefficients are related or coincide with previously introduced entanglement witnesses, as indicated by the references. For a hierarchical ordering of the coefficients, see Fig.~\ref{fig:hierarchy}.} %All of them can be derived from the general criterion introduced in Ref.~\cite{Gessner2016} using different constraints and approximations.}
\label{tab:coeffs}
\end{table*}

\begin{figure}[tb]
\includegraphics[width=.4\textwidth]{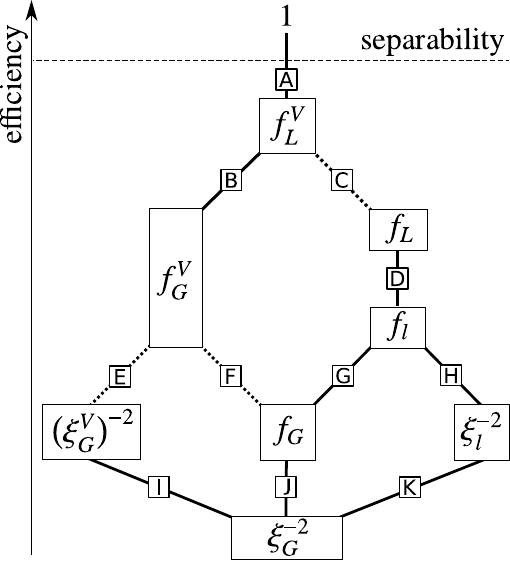}
\caption{Schematic hierarchy of locally and globally optimized coefficients based on the Fisher information or spin squeezing, with or without comparison to local variances. For separable states, the values of all these coefficients are bounded by $1$ (dashed horizontal line). The solid lines connect a more efficient entanglement coefficient (above) to a lower bound (below). The dotted lines indicate that all entangled states that are detected by the less efficient coefficient (below) will also be detected by the more efficient one (above), while they may not always provide quantitative bounds for each other. The letters allow identification of the relation with their statement in the text: A,~Eq.~(\ref{eq:sepcriterion}); B,~Eq.~(\ref{eq:flvfgv}); C,~Eq.~(\ref{eq:flnflv}); D,~Eq.~(\ref{eq:fLfl}); E,~Eq.~(\ref{eq:xigvfgv}); F,~Eq.~(\ref{eq:fgnfgv}); G,~Eq.~(\ref{eq:flnfgn}); H,~Eq.~(\ref{eq:flnxil}); I,~Eq.~(\ref{eq:xigvxign}); J,~Eq.~(\ref{eq:fgnxign}); and K,~Eq.~(\ref{eq:xilxign}).}
\label{fig:hierarchy}
\end{figure}

Above we introduced a series of separability criteria based on the Fisher densities [see Eqs.~(\ref{eq:flv}), (\ref{eq:fln}), (\ref{eq:fgv}), and~(\ref{eq:fgn})] or its lower bounds leading to spin-squeezing coefficients [see Eqs.~(\ref{eq:xil}), (\ref{eq:xign}), and~(\ref{eq:xigv})]. An overview is further given in Table~\ref{tab:coeffs} together with general requirements for the calculation of the coefficients. In this section we compare the different coefficients. Their relationships are summarized in Fig.~\ref{fig:hierarchy}. There, the higher coefficients provide more stringent separability criteria.

The most powerful separability criterion is provided by $f_L^V\leq 1$, where $f_L^V$ is defined in Eq.~(\ref{eq:flv}). This criterion reveals the entanglement of arbitrary pure states. To find the optimal vector for $f_L^V$ all elementary spin-$1/2$ systems are optimized locally. By imposing the additional constraint that all local vectors must coincide, we obtain the globally optimized, variance-assisted Fisher density $f^V_G$, Eq.~(\ref{eq:fgv}), hence providing a lower bound on $f_L^V$:
\begin{align}\label{eq:flvfgv}
f_L^V\geq f^V_G.
\end{align}
This and all the other relations reported below are valid for an arbitrary state $\hat{\rho}$. By the same argument we also find
\begin{align}\label{eq:flnfgn}
f_l\geq f_G,
\end{align}
where $f_l$ and $f_G$ were introduced in Eqs.~(\ref{eq:flnn}) and (\ref{eq:fgn}), respectively, as well as
\begin{align}\label{eq:xilxign}
\xi_l^{-2}\geq \xi_G^{-2},
\end{align}
introduced in Eqs.~(\ref{eq:xil}) and (\ref{eq:xign}). Furthermore, the additional local normalization constraints that are required for $f_l$ but not for $f_L$ lead to
\begin{align}\label{eq:fLfl}
f_L \geq f_l.
\end{align}

The locally optimized Fisher density $f_l$ can be further bounded by the locally optimized spin-squeezing coefficient $\xi_l$, defined in Eq.~(\ref{eq:xil}):
\begin{align}\label{eq:flnxil}
f_l\geq \xi_l^{-2}.
\end{align}
To see this, let $\mathbf{c}_{1}$ denote the locally normalized vector that achieves the maximum in Eq.~(\ref{eq:flnn}). We find
\begin{align}
f_l(\hat{\rho})&=\frac{F_Q[\hat{\rho},\hat{A}(\mathbf{c}_{1})]}{N}\notag\\
%&\geq \max_{\hat{B}}\frac{|\langle [\hat{A}(\mathbf{c}_{1}),\hat{B}]\rangle_{\hat{\rho}}|^2}{N\mathrm{Var}(\hat{B})_{\hat{\rho}}}\notag\\
&\geq \max_{\mathbf{c}'}\frac{|\langle [\hat{A}(\mathbf{c}_{1}),\hat{A}(\mathbf{c}')]\rangle_{\hat{\rho}}|^2}{N\mathrm{Var}(\hat{A}(\mathbf{c}'))_{\hat{\rho}}}\notag\\
&\geq \max_{\mathbf{c}}\frac{\langle \hat{A}(\mathbf{c}_0)\rangle_{\hat{\rho}}^2}{N\mathrm{Var}(\hat{A}(\mathbf{c}))_{\hat{\rho}}}\label{eq:step1}\\
&=\frac{1}{\xi_l^{2}(\hat{\rho})},
\end{align}
where in the first step we have used Eq.~(\ref{eq:FBound}) and the maximization in~(\ref{eq:step1}) is performed over vectors $\mathbf{c}$ that are locally normalized and locally orthogonal to the local mean spin $\mathbf{c}_0$. Following the same steps for a global optimization over $\mathbf{n}\in \mathbb{R}^3$, one finds
\begin{align}\label{eq:fgnxign}
f_G\geq\xi_G^{-2}.
\end{align}
This relationship is well known \cite{LucaRMP} and shows that the Wineland spin-squeezing coefficient $\xi_G$ \cite{PhysRevA.46.R6797,PhysRevA.50.67}, Eq.~(\ref{eq:xign}), is an entanglement witness \cite{Sorensen2001}, but the Fisher density $f_G$, Eq.~(\ref{eq:fgn}), detects entanglement 
more efficiently \cite{PhysRevLett.102.100401}.

The relation between Eqs.~(\ref{eq:xign}) and (\ref{eq:xigv}) is given by
\begin{align}\label{eq:xigvxign}
(\xi^V_G)^{-2}\geq \xi_G^{-2},
\end{align}
as a direct consequence of the upper bound for the local variances~(\ref{eq:upperboundLclVar}).

In some cases, we cannot establish an inequality between the coefficients due to the different optimizations involved in their definitions. Yet the violation of one separability criterion may imply the violation of another. For instance, $f^V_G$ provides a stronger entanglement criterion than $f_G$, i.e., 
\begin{align}\label{eq:fgnfgv}
f_G>1\quad\Rightarrow\quad f^V_G>1.
\end{align}
To prove this, we notice that from $f_G(\hat{\rho})>1$ it follows that there exists one vector $\mathbf{n}_{\mathrm{opt}}$ for which $F_Q[\hat{\rho},\hat{J}_{\mathbf{n}_{\mathrm{opt}}}]>N$. Now let $\mathbf{n}_{\max}$ denote the vector that maximizes the quantity $F_Q[\hat{\rho},\hat{J}_{\mathbf{n}_{\max}}]-4\mathrm{Var}(\hat{J}_{\mathbf{n}_{\max}})_{\Pi(\hat{\rho})}$. We find that $F_Q[\hat{\rho},\hat{J}_{\mathbf{n}_{\max}}]-4\mathrm{Var}(\hat{J}_{\mathbf{n}_{\max}})_{\Pi(\hat{\rho})}\geq F_Q[\hat{\rho},\hat{J}_{\mathbf{n}_{\mathrm{opt}}}]-4\mathrm{Var}(\hat{J}_{\mathbf{n}_{\mathrm{opt}}})_{\Pi(\hat{\rho})}\geq F_Q[\hat{\rho},\hat{J}_{\mathbf{n}_{\mathrm{opt}}}]-N > 0$, and hence, $f^V_G(\hat{\rho})>1$, where we used Eq.~(\ref{eq:upperboundLclVar}). 

Analogously, we find $F_Q[\hat{\rho},\hat{A}(\mathbf{c}_{\max})]-4\mathrm{Var}(\hat{A}(\mathbf{c}_{\max}))_{\Pi(\hat{\rho})}\geq F_Q[\hat{\rho},\hat{A}(\mathbf{c}_{\mathrm{opt}})]-N > 0$, and thus,
\begin{align}\label{eq:flnflv}
f_L>1\quad\Rightarrow\quad f^V_L>1.
\end{align}

Finally, we show that the variance-assisted Fisher density $f^V_G$ is more efficient than the variance-assisted spin-squeezing coefficient $\xi^V_G$, i.e.,
\begin{align}\label{eq:xigvfgv}
(\xi^V_G)^{-2}>1\quad\Rightarrow\quad f^V_G>1.
\end{align}
For the proof we first we note that $(\xi^V_G)^{-2}(\hat{\rho})>1$ implies that $\langle J_{\mathbf{n}_0}\rangle_{\hat{\rho}}^2/\mathrm{Var}(J_{\mathbf{n}^{\max}_{\perp}})_{\hat{\rho}}>4\mathrm{Var}(\hat{J}_{\mathbf{n}'_{\perp}})_{\Pi(\hat{\rho})}$ for some $\mathbf{n}^{\mathrm{max}}_{\perp}$; recall Eq.~(\ref{eq:xigv}). By construction, the vector $\mathbf{n}^{\mathrm{max}}_{\perp}$ forms a mutually orthonormal basis of $\mathbb{R}^3$ together with the mean-spin vector $\mathbf{n}_0$, defined in Eq.~(\ref{eq:globalmeanspin}), and a third vector $\mathbf{n}'_{\perp}$. For $\mathbf{n}_{\max}$ defined as before, we now obtain
\begin{align}
F_Q[\hat{\rho},\hat{J}_{\mathbf{n}_{\max}}]-4\mathrm{Var}(\hat{J}_{\mathbf{n}_{\max}})_{\Pi(\hat{\rho})}&\geq F_Q[\hat{\rho},\hat{J}_{\mathbf{n}'_{\perp}}]-4\mathrm{Var}(\hat{J}_{\mathbf{n}'_{\perp}})_{\Pi(\hat{\rho})}\notag\\&\geq \frac{|\langle [J_{\mathbf{n}'_{\perp}},J_{\mathbf{n}_{\perp}}]\rangle_{\hat{\rho}}|^2}{\mathrm{Var}(J_{\mathbf{n}^{\max}_{\perp}})_{\hat{\rho}}}-4\mathrm{Var}(\hat{J}_{\mathbf{n}'_{\perp}})_{\Pi(\hat{\rho})}\notag\\&\geq \frac{\langle J_{\mathbf{n}_0}\rangle_{\hat{\rho}}^2}{\mathrm{Var}(J_{\mathbf{n}^{\max}_{\perp}})_{\hat{\rho}}}-4\mathrm{Var}(\hat{J}_{\mathbf{n}'_{\perp}})_{\Pi(\hat{\rho})}\notag\\&>0,
\end{align}
where we used Eq.~(\ref{eq:FBound}). From this we follow $f^V_G(\hat{\rho})>1$, as claimed.

\subsection{Relation of the coefficients to metrological quantum gain}\label{sec:metro}
The Fisher information plays a central role in quantum metrology \cite{LucaRMP,PhysRevLett.96.010401,paris2009,Giovannetti2011,Varenna,Helstrom1976}. Let us consider the transformation of a probe state $\hat{\rho}$ by the unitary operator $e^{-i\theta\hat{A}(\mathbf{c})}$. This transformation may assign different weights and directions to the individual qubits. The ultimate precision for the estimation of the phase $\theta$ is given by the quantum Cram\'{e}r-Rao bound 
\begin{align}
\mathrm{Var}(\theta_{\mathrm{est}})_{\hat{\rho}}\geq\frac{1}{F_Q[\hat{\rho},\hat{A}(\mathbf{c})]},
\end{align}
where $\theta_{\mathrm{est}}$ is an arbitrary, unbiased estimator of the phase $\theta$. This bound is saturable by optimal measurements \cite{PhysRevLett.72.3439}. According to Eq.~(\ref{eq:sepcriterion}), for a separable state $\hat{\rho}_{\mathrm{sep}}$, we have
\begin{align}
\mathrm{Var}(\theta_{\mathrm{est}})_{\hat{\rho}_{\mathrm{sep}}} \geq \frac{1}{4\sum_{i=1}^N \mathrm{Var}\big(\hat{A}_i\big)_{\hat{\rho}_{\mathrm{sep}}}}
\end{align}
Taking the highest possible value of the sum of local variances, Eq.~(\ref{eq:upperboundLclVar}), we obtain $\mathrm{Var}(\theta_{\mathrm{est}})_{\hat{\rho}_{\mathrm{sep}}}\geq \mathrm{Var}(\theta_{\mathrm{est}})_{\textrm{SN}}$, where
\begin{align}\label{eq:generalshotnoise}
\mathrm{Var}(\theta_{\mathrm{est}})_{\mathrm{SN}}=\frac{1}{|\mathbf{c}|^2}
\end{align}
represents the highest value of phase sensitivity that can be reached by separable states and thus generalizes the notion of the 
shot-noise (or standard quantum) limit \cite{LucaRMP} to the case of inhomogeneous probing \cite{PezzePNAS2016}. 
In the homogeneous case, $|\mathbf{n}_i|^2=1$ (and thus $|\mathbf{c}|^2=\sum_{i=1}^N|\mathbf{n}_i|^2=N$), Eq.~(\ref{eq:generalshotnoise}) reduces to the usual 
definition of shot noise, $\mathrm{Var}(\theta_{\mathrm{est}})_{\mathrm{SN}}=1/N$ \cite{PhysRevLett.102.100401,PhysRevLett.96.010401}. %Hence, any value above this limit indicates a metrological quantum advantage over the best possible classical strategy. The highest achievable precision when taking entangled quantum states as resource into account, is the so-called Heisenberg limit $\mathrm{Var}(\theta_{\mathrm{est}})_{\hat{\rho}}= 1/N$ \cite{LucaRMP,PhysRevLett.102.100401,PhysRevLett.96.010401}.

The coefficient $f_l$ and all of its lower bounds $f_G$, $\xi_G$, and $\xi_l$ are therefore suitable quantifiers of the metrological quantum gain in a system of a fixed particle number $N$. This is also true for $f_L$; however, for a quantitative interpretation one must keep in mind that only a global normalization constraint on the generator of the unitary evolution is required. As will be emphasized by an example in the next section, the absence of local normalization constraints can lead to an effective amplification of some subsystems which is not possible under the conditions considered for the other coefficients. Furthermore, notice that all of the introduced entanglement coefficients coincide at the value $1$ for a spin-coherent state, which is often used as a benchmark for the optimal precision of separable quantum states \cite{PhysRevA.46.R6797}.

All variance-based coefficients (superscript $V$) are constructed in a way that optimizes their ability to identify entangled quantum states. They compare the value of the Fisher information to a state-dependent separability bound, and their value is therefore not correlated with the quantum gain of this particular quantum state over all possible separable states.

\subsection{Relation of the coefficients to multipartite entanglement}\label{sec:multi}
Let us first recall some basic definitions \cite{Guhne2009,PhysRevA.65.012107}. An $N$-partite pure state $|\Psi_{\mathrm{k-prod}}\rangle$ is called $k$-producible if it can be written as a product state of local states that do not contain more than $k$-particle entanglement, i.e., the state can be decomposed as
\begin{align}
|\Psi_{\mathrm{k-prod}}\rangle=\bigotimes_{l=1}^M|\varphi_l\rangle,
\end{align}
where $|\varphi_l\rangle$ describes a quantum state of $N_l\leq k$ particles with $\sum_{l=1}^MN_l=N$. A density matrix is called $k$-producible if it can be written as a convex combination of $k$-producible pure states, i.e.,
\begin{align}
\hat{\rho}_{\mathrm{k-prod}}=\sum_{\gamma}p_{\gamma}|\Psi^{\gamma}_{\mathrm{k-prod}}\rangle\langle\Psi^{\gamma}_{\mathrm{k-prod}}|.
\end{align}
Conversely, a state is called $k$-partite entangled if it is $k$-producible but not $(k-1)$-producible.

It was shown that any $k$-producible state must satisfy \cite{PhysRevA.85.022321,PhysRevA.85.022322}
\begin{align}\label{eq:ksepbound}
f_l(\hat{\rho}_{\mathrm{k-prod}})\leq \frac{sk^2 + r}{N},
\end{align}
where $s=\lfloor N/k\rfloor$ is the largest integer $\leq N/k$ and $r=N-sk$. If $N/k$ is an integer, the right-hand side in Eq.~(\ref{eq:ksepbound}) reduces to $k$. A similar relation is presently not available for $f_L$ where the local vectors are not subject to normalization or for all coefficients that involve measurements of the local variances.

By providing lower bounds to $f_l$, the quantities $f_G$, $\xi_G$, and $\xi_l$ must all respect the same bound~(\ref{eq:ksepbound}) for arbitrary $k$-separable quantum states. Hence, we may use, e.g., the local spin-squeezing coefficient $\xi_l$ to quantify multipartite entanglement without measurements of the Fisher information. For the detection of multipartite entanglement from global spin-squeezing parameters, such as  $\xi_G$, see \cite{PhysRevLett.86.4431,PhysRevA.86.012337}.

In summary, we can classify the coefficients presented here into two categories. The coefficients $f_G$, $f_l$, $\xi_G$, and $\xi_l$ are quantitatively meaningful in a sense that they assess the metrological quantum gain and the multipartite entanglement of the state in question. The variance-assisted coefficients $f^V_l$, $f^V_G$, and $\xi^V_G$ are more efficient at detecting bipartite entanglement; however, their value is not quantitatively meaningful. The coefficient $f_L$ represents a special case, as it is quantitatively meaningful for metrology but not able to identify multipartite entanglement.

\subsection{Examples}\label{sec:qubitexample1}
We now illustrate the differences between the coefficients with a series of examples. The examples focus on the enhancement due to local optimizations, as well as the role of local normalization constraints. For an example that highlights the relevance of the local variances, we refer to Ref.~\cite{Gessner2016}.

\subsubsection{Local normalization constraints}
The example below reveals how the absence of local normalization constraints allows us to assign more weight to the entangled part of the state, thereby enhancing the efficiency of the entanglement detection. Let us consider the following state:
\begin{align}\label{eq:example1}
\hat{\rho}_{N,K}=|\mathrm{GHZ}_{K}\rangle\langle \mathrm{GHZ}_{K}|\otimes\frac{\hat{\mathbb{I}}_{(N-K)}}{(N-K)^2},
\end{align}
where $|\mathrm{GHZ}_N\rangle=(|\uparrow_z\rangle^{\otimes N}+|\downarrow_z\rangle^{\otimes N})/\sqrt{2}$ denotes a Greenberger-Horne-Zeilinger (GHZ) state of $N$ particles, $|\uparrow_z\rangle$ and $|\downarrow_z\rangle$ are eigenstates of $\hat{\sigma}^{(z)}$, and $\hat{\mathbb{I}}_{N}$ is the identity operator on $(\mathbb{C}^2)^{\otimes N}$. %This state achieves the Heisenberg limit in the first $K$ subsystems, whereas no phase sensitivity at all can be achieved in the remaining $N-K$ subsystems. 
Local and globally optimized coefficients yield the same results due to the permutation symmetry within the GHZ state and the absence of a preferred direction for the identity. All single-qubit reduced density matrices of this state are maximally mixed, and therefore, the variance-assisted coefficients yield the same values as those that use the upper bound. From the additivity of the Fisher information \cite{Varenna} it follows immediately that
\begin{align}
f_l(\hat{\rho}_{N,K})=f_G(\hat{\rho}_{N,K})=\frac{K^2}{N},
\end{align}
where the maximum is obtained when all local directions coincide at $\mathbf{n}_i=\mathbf{e}_z$ and $\mathbf{e}_z\in\mathbb{R}^3$ is a unit vector along $z$. Consequently, for $K<\sqrt{N}$ the entanglement of the state $\hat{\rho}_{N,K}$ is no longer detected by the coefficients $f_l$ and $f_G$. The locally unconstrained optimization of the coefficient $f_L$, however, opens up the possibility to effectively ignore the incoherent $N-K$ subsystems by setting $\mathbf{n}_i=0$ for $i=N-K+1,\dots,N$ %The global normalization condition $|\mathbf{c}|^2=N$ then leads to an amplification of the phase signal in the remaining $K$ subsystems, which are highly coherent. 
%To satisfy normalization, the nonzero elements of $\mathbf{c}$ must be enhanced by a factor of $\sqrt{N/K}$. In other words, the local optimization allows 
and to assign all the weight to the maximally entangled GHZ state. The maximum value
\begin{align}\label{eq:flrhok}
f_L(\hat{\rho}_{N,K})=K
\end{align}
is attained when $\mathbf{n}_i=\sqrt{N/K}\mathbf{e}_z$ for $i=1,\dots,K$. Hence, the coefficient $f_L$ reveals that the state $\hat{\rho}_{N,K}$ is always entangled for $K>1$. The result~(\ref{eq:flrhok}) indeed reflects that for the $K$-qubit GHZ state $f_l(|\mathrm{GHZ}_{K}\rangle)=f_G(|\mathrm{GHZ}_{K}\rangle)=K$.

\subsubsection{Local vs global optimization}
We now discuss some examples that highlight the relevance of local manipulations for the detection of entanglement. First, we consider the following set of ``twisted'' GHZ states of $N=3K$ particles:
\begin{align}
|\mathrm{GHZ}^t_{3K}\rangle&=\frac{1}{\sqrt{2}}\left(|\uparrow_x\rangle^{\otimes K}\otimes|\uparrow_y\rangle^{\otimes K}\otimes|\uparrow_z\rangle^{\otimes K}\right.\notag\\&\qquad+\left.|\downarrow_x\rangle^{\otimes K}\otimes|\downarrow_y\rangle^{\otimes K}\otimes|\downarrow_z\rangle^{\otimes K}\right),
\end{align}
where we introduced the eigenvectors of the three Pauli matrices as $\hat{\sigma}^{(\alpha)}=|\uparrow_{\alpha}\rangle\langle\uparrow_{\alpha}|-|\downarrow_{\alpha}\rangle\langle\downarrow_{\alpha}|$ for $\alpha=x,y,z$. 

By construction, these states are highly asymmetric, and therefore, we expect entanglement detection strategies that allow for flexible, individual tuning of the local constituents to be advantageous over global methods. We introduce the locally optimized, twisted linear operator $\hat{A}(\mathbf{c}^t)$ [recall Eq.~(\ref{eq:AofC})], where the constituents of $\mathbf{c}^t=(\mathbf{n}^t_1,\dots,\mathbf{n}^t_N)$ are chosen as $\mathbf{n}^t_1=\cdots=\mathbf{n}^t_K=\mathbf{e}_x$, $\mathbf{n}^t_{K+1}=\cdots=\mathbf{n}^t_{2K}=\mathbf{e}_y$, and $\mathbf{n}^t_{2K+1}=\cdots=\mathbf{n}^t_N=\mathbf{e}_z$. A straightforward calculation now reveals that the quantum Fisher information attains its largest possible value,
\begin{align}
F_Q[|\mathrm{GHZ}^t_{3K}\rangle,\hat{A}(\mathbf{c}^t)]=4\langle\mathrm{GHZ}^t_{3K}|\hat{A}^2(\mathbf{c}^t)|\mathrm{GHZ}^t_{3K}\rangle=N^2.
\end{align}
We immediately see that $f_l(|\mathrm{GHZ}^t_{3K}\rangle)=N$. As discussed in Secs.~\ref{sec:metro} and~\ref{sec:multi}, this indicates $N$-partite entanglement, i.e., genuine multipartite entanglement in the state $|\mathrm{GHZ}^t_{3K}\rangle$, as well as a maximal metrological quantum gain in a suitable interferometer.

\begin{figure}
\centering
\includegraphics[width=.45\textwidth]{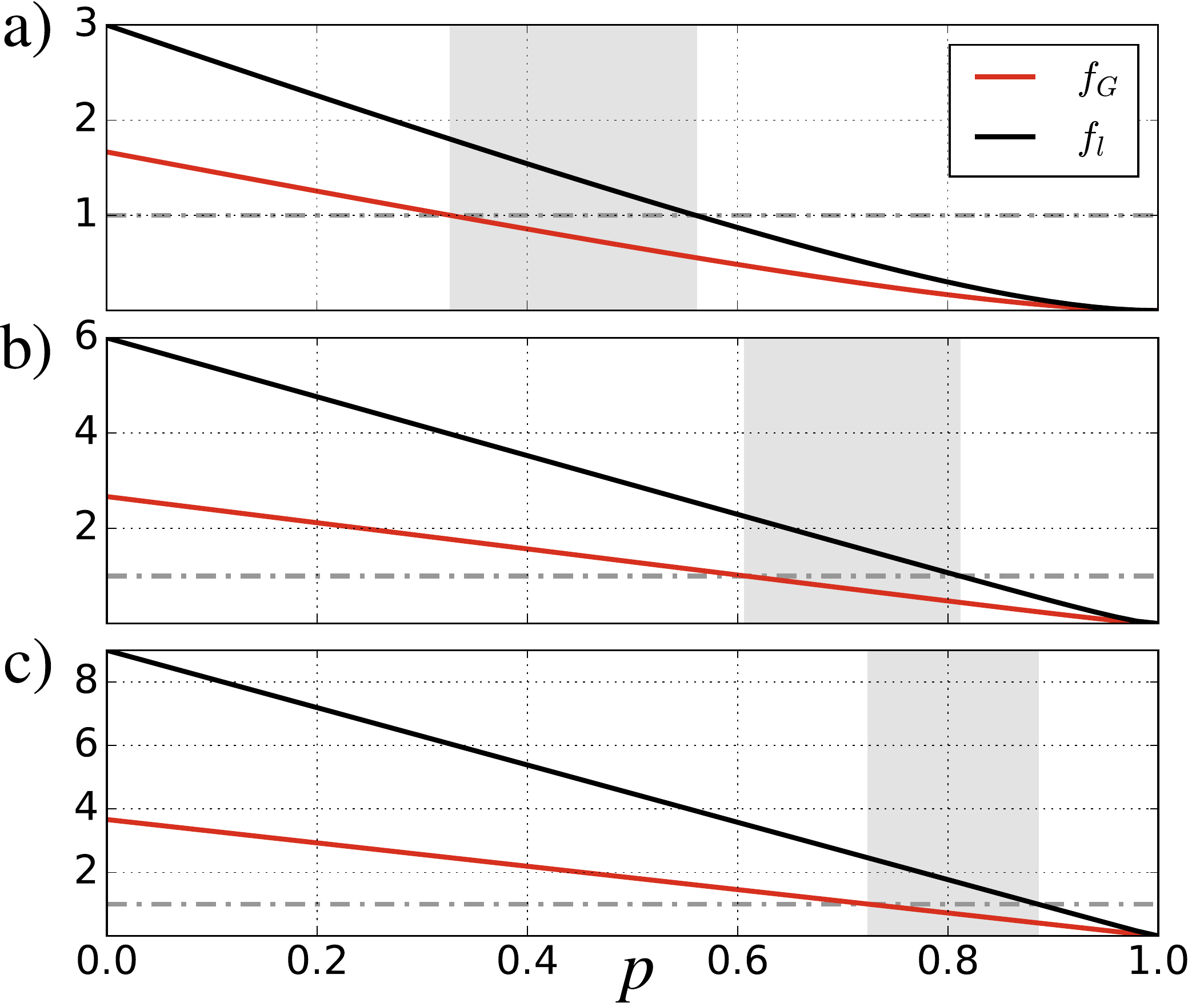}
\caption{(Color online) Coefficients $f_l$ and $f_G$ for the noisy ``twisted'' GHZ states $\hat{\rho}^t(p)$, Eq.~(\ref{eq:noisytwistedGHZ}) with $N=3$ (a), $N=6$ (b), and $N=9$ (c). In the gray shaded areas (where $f_l>1$ while $f_G\leq 1$), the entanglement of the state can only be revealed by local access to the parties. The other coefficients either coincide with those already plotted, i.e., $f_L\equiv f_L^V\equiv f_l$ and $f_G^V\equiv f_G$, or vanish $\xi_l^{-2}\equiv \xi_G^{-2}\equiv (\xi_G^V)^{-2}\equiv 0$ for all $p$.}
\label{fig:LvsGQFI}
\end{figure}

This result can be compared to the one obtained from a global optimization as in $f_G$, Eq.~(\ref{eq:fgn}). We thus consider the global quantum Fisher matrix, Eq.~(\ref{eq:globalQFM}), leading to
\begin{align}
\mathbf{Q}^{G}_{|\mathrm{GHZ}^t_{3K}\rangle}=\begin{pmatrix}
K(K+2) & K^2 & K^2 \\
K^2 & K(K+2) & K^2 \\
K^2 & K^2 & K(K+2)
\end{pmatrix}.
\end{align}
The maximum eigenvalue of $2K+3K^2$ is obtained along the direction $\mathbf{n}^t=(1,1,1)/\sqrt{3}$. Hence, 
\begin{align}
f_G(|\mathrm{GHZ}^t_{3K}\rangle)=\frac{F_Q[|\mathrm{GHZ}^t_{3K}\rangle,\hat{J}_{\mathbf{n}^t}]}{3K}=K+\frac{2}{3},
\end{align}
and thus, the global coefficient $f_G$ is able to reveal only $K$-partite entanglement of this state and expresses a lower quantum gain in a collective interferometer compared to the locally optimized scenario above.

In the presence of white noise, the local access to the state becomes crucial even to reveal its inseparability. To show this, we compare the global and local Fisher densities $f_G$ and $f_l$ for mixtures of the twisted GHZ state with the maximally mixed state, 
\begin{align}\label{eq:noisytwistedGHZ}
\hat{\rho}^t(p)=\frac{1}{1+p}\left(|\mathrm{GHZ}^t_{3K}\rangle\langle \mathrm{GHZ}^t_{3K}|+p\frac{\hat{\mathbb{I}}_{3K}}{(3K)^2}\right).
\end{align}
The plot in Fig.~\ref{fig:LvsGQFI} reveals a finite parameter range in which the entanglement of $\hat{\rho}^t(p)$ is detected only by the local Fisher density, whereas its global counterpart is not able to achieve this.

%We do not discuss an example to highlight the role of the local variances, since such a family of $N$-qubit states was already presented in Ref.~\cite{Gessner2016}. 

\begin{figure}[tb]
\centering
\includegraphics[width=.45\textwidth]{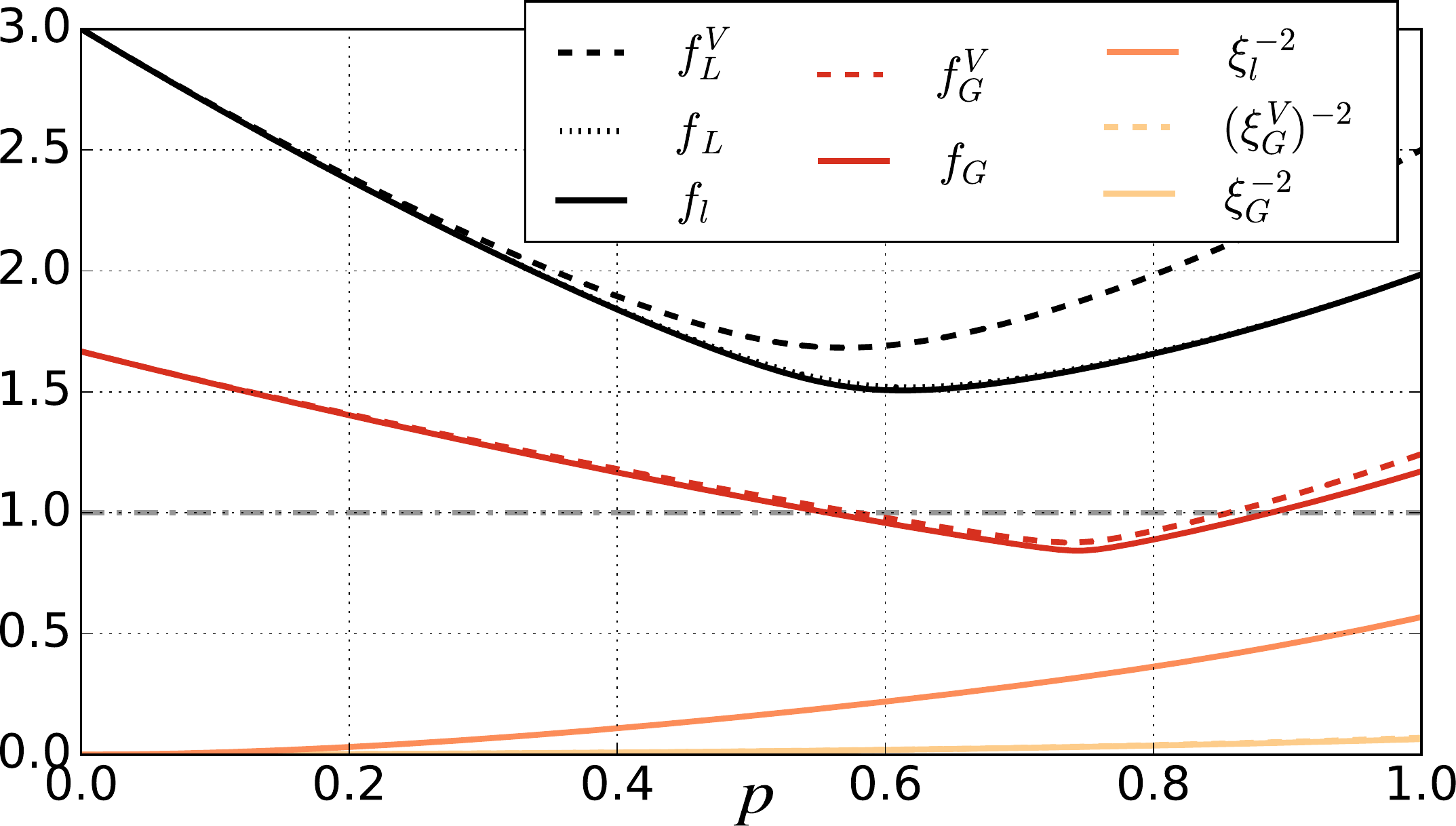}
\caption{Entanglement coefficients for the states~(\ref{eq:twisted}) for $N=3$ as a function of $p$. The squeezing coefficients never exceed the separability limit $1$. In contrast, the Fisher densities are able to reveal the state's entanglement. However, local access is required in a finite range of $0.6\lesssim p\lesssim 0.9$.}
\label{fig:twistedWGHZ}
\end{figure}

\subsubsection{Mixture of twisted W and GHZ states}
To apply our criteria to a slightly more complex family of example states, we introduce the following type of mixture:
\begin{align}\label{eq:twisted}
\hat{\rho}_{3K}^t(p)=\left(p|\mathrm{GHZ}^t_{3K}\rangle\langle \mathrm{GHZ}^t_{3K}|+(1-p)|\mathrm{W}^t_{3K}\rangle\langle \mathrm{W}^t_{3K}|\right).
\end{align}
where $0\leq p\leq 1$ and
\begin{align}
|\mathrm{W}^t_{3K}\rangle&=\frac{1}{\sqrt{3}}\left(|\uparrow_x\rangle^{\otimes K}\otimes|\downarrow_x\rangle^{\otimes K}\otimes|\downarrow_x\rangle^{\otimes K}\right.\notag\\&\qquad+|\downarrow_y\rangle^{\otimes K}\otimes|\uparrow_y\rangle^{\otimes K}\otimes|\downarrow_y\rangle^{\otimes K}\notag\\&\qquad+\left.|\downarrow_z\rangle^{\otimes K}\otimes|\downarrow_z\rangle^{\otimes K}\otimes|\uparrow_z\rangle^{\otimes K}\right)
\end{align}
is a twisted W state of $3K$ particles. As displayed in Fig.~\ref{fig:twistedWGHZ}, the contribution of the state $|\mathrm{W}^t_{3K}\rangle$ leads to a significant improvement of the entanglement detection by means of the local variances. While the spin-squeezing coefficients $\xi_l^{-2}$, $\xi_G^{-2}$ $(\xi_G^V)^{-2}$ are no longer zero when $p>0$, they never exceed the separability threshold of $1$ and hence are unable to detect the entanglement. The state is successfully identified as entangled by the Fisher densities. We observe a strong advantage of locally optimized criteria over global ones, including a finite parameter range of $p$ in which only the locally optimized Fisher densities are able to reveal the entanglement.

\subsubsection{Enhanced entanglement detection in quantum simulators}
We now turn to a dynamical example, generated by the long-range Ising Hamiltonian with transverse field \cite{PorrasCiracSpins,PhysRevB.93.155153,PhysRevA.87.042101,Richerme2014,Jurcevic2014,Bohnet1297},
\begin{align}\label{eq:longrangeH}
H_{\alpha,B} =\frac{1}{N}\sum_{i>j}^NJ_{ij}\hat{\sigma}^{(z)}_i\hat{\sigma}^{(z)}_i + B\sum_{i=1}^N\hat{\sigma}^{(x)}_i,
\end{align}
where $J_{ij}=J_0/|i-j|^{\alpha}$ determines the spin-spin interaction strength $J_0$ and range $\alpha$, and $B$ denotes the strength of the transverse magnetic field. This model can be realized in trapped-ion quantum simulators \cite{PorrasCiracSpins,Richerme2014,Jurcevic2014,Bohnet1297} with approximately $0\leq \alpha \leq 3$; however, smaller values of $\alpha$ are more easily accessible \cite{Bohnet1297}. In the special case $\alpha=0$, this reduces to the Lipkin-Meshkov-Glick model \cite{LMG}. Its dynamics has been frequently employed to generate spin-squeezed and non-Gaussian entangled quantum states \cite{LucaRMP}. For $B=0$ this Hamiltonian is also known as \textit{one-axis twisting} \cite{SpinSqueezing,Esteve2008,LucaRMP}.

\begin{figure}[tb]
\centering
\includegraphics[width=.45\textwidth]{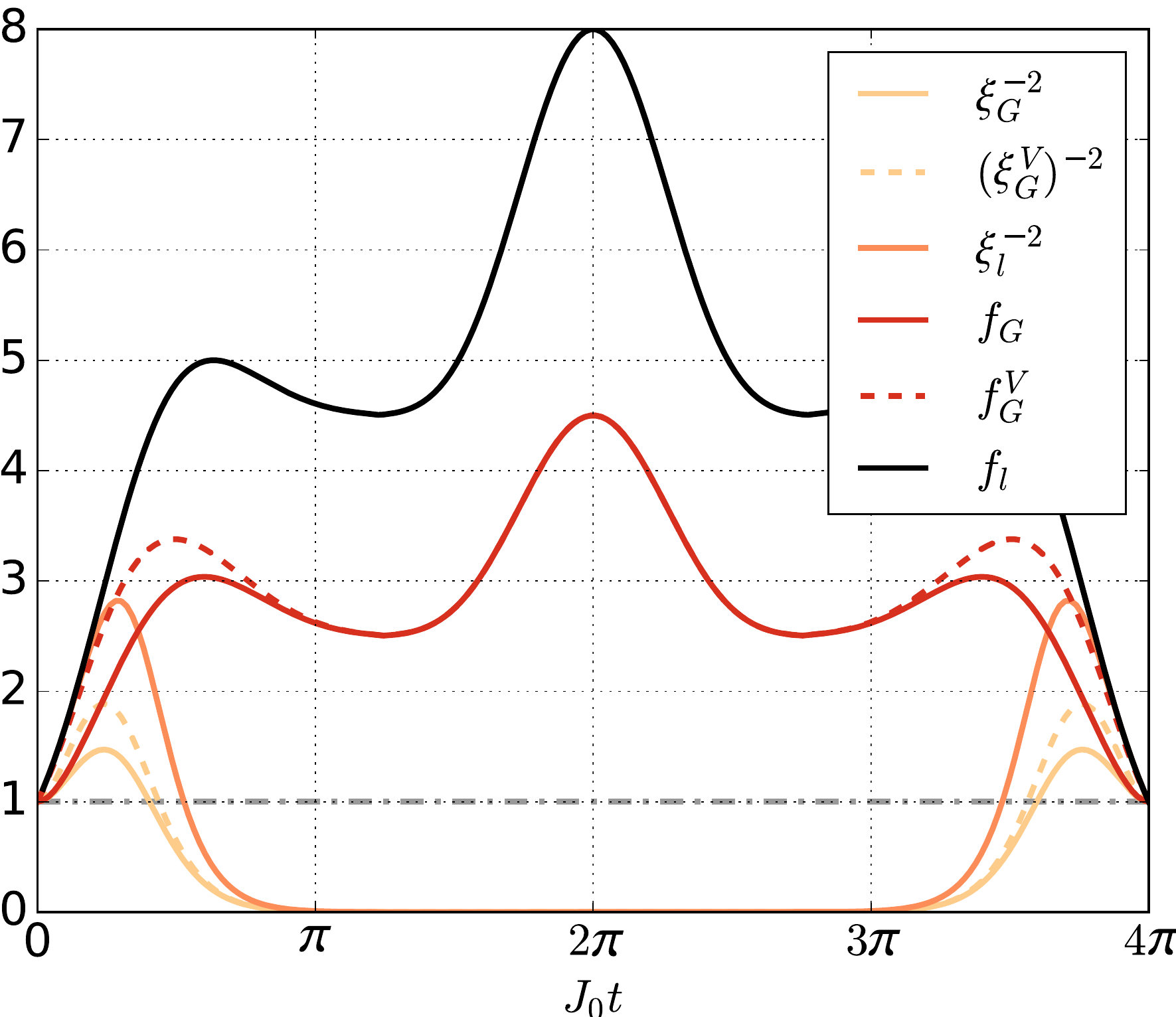}
\caption{Entanglement coefficients for the one-axis twisting dynamics, governed by $H_{0,0}$ [Eq.~(\ref{eq:longrangeH})] with an asymmetric initial state~(\ref{eq:asyminit}) and $N=8$ spins. We have $f_L\equiv f_L^V\equiv f_l$ at all times but only the latter is shown in the plot. The horizontal dash-dotted line at value $1$ indicates the separability bound.}
\label{fig:asym1axtwist}
\end{figure}

To generate an intrinsically asymmetric situation, we assume the following separable initial state of $N=2M$ qubits:
\begin{align}\label{eq:asyminit}
|\Psi_0\rangle=|\downarrow_y\rangle^{\otimes M}\otimes|\downarrow_x\rangle^{\otimes M}.
\end{align}
The behavior of the various entanglement coefficients is shown in Fig.~\ref{fig:asym1axtwist} for the evolution of $|\Psi_0\rangle$ under the one-axis twisting evolution, i.e., $|\Psi(t)\rangle=e^{-iH_{0,0}t}|\Psi_0\rangle$. By restricting ourselves to global observables as in $f_G$ and $\xi_G$ the entanglement of the states is only partially or not at all revealed. At initial times, a slight enhancement due to local variances can be observed. At long times, the spin-squeezing coefficients are no longer able to detect entanglement due to the increasingly non-Gaussian nature of the state. At $J_0t=2\pi$ the state reaches the maximal possible value of $f_l=N$. The local spin-squeezing criterion $\xi_l$ outperforms the global ones $\xi_G$ and $\xi_G^V$ and, at short times, even the global Fisher density $f_G$.

\begin{figure}[tb]
\centering
\includegraphics[width=.45\textwidth]{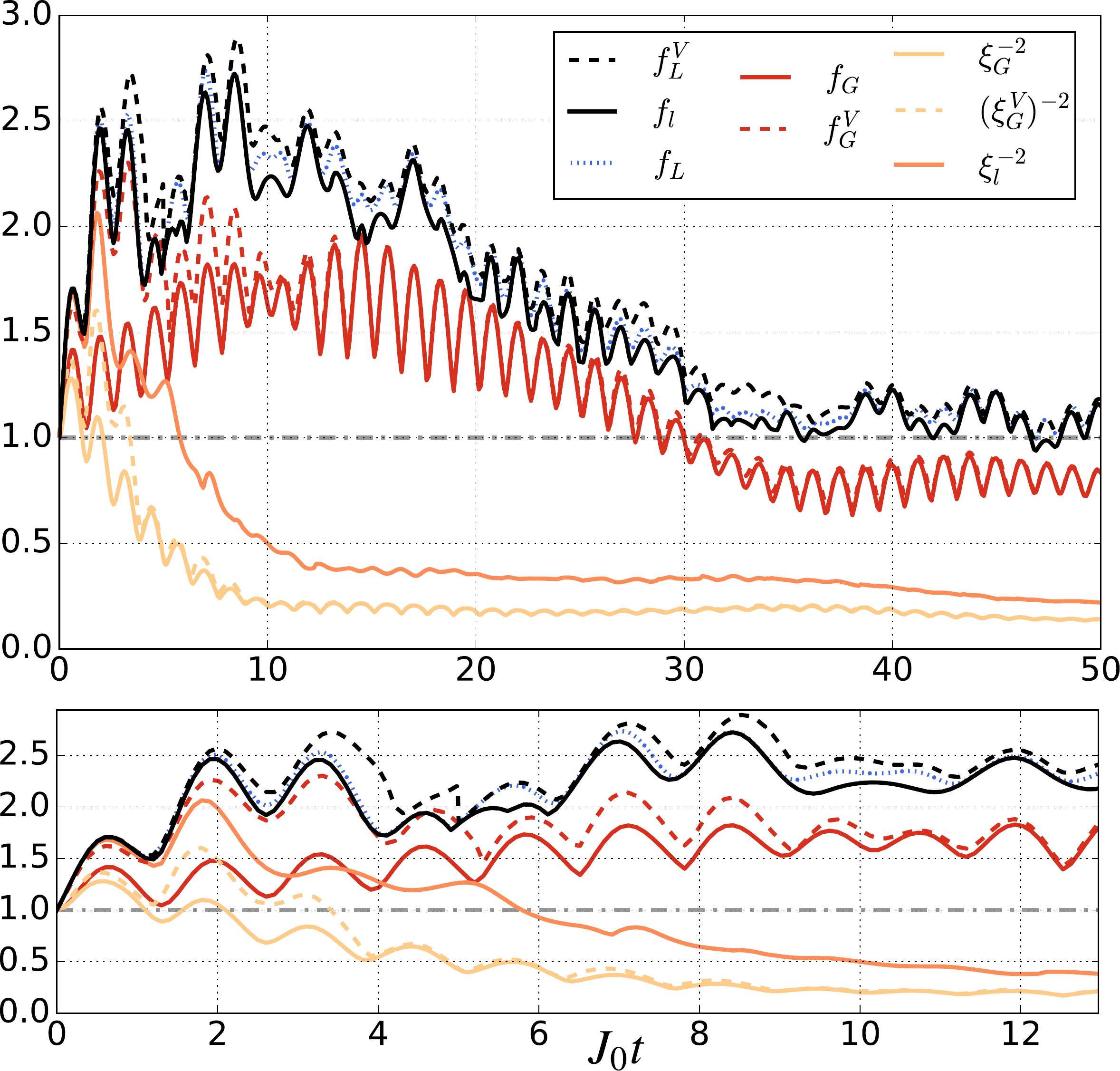}
\caption{Evolution of the entanglement coefficients for the Ising model with transverse field $B=J_0$ and long-range interactions $\alpha=0.2$, including realistic noise processes described by Eq.~(\ref{eq:noisydyn}), for $N=8$ spins. The separability limit is indicated by a dashed-dotted line at value $1$. The lower panel shows a magnified display of the evolution at initial times. The complex dynamics close to the quantum phase transition together with the asymmetric initial condition~(\ref{eq:asyminit}) opens up the full spectrum of entanglement coefficients, summarized in Tab.~\ref{tab:coeffs}. The relations displayed in Fig.~\ref{fig:hierarchy} can be observed---in particular the advantage of local coefficients over global ones. Notice also that at early times the local spin squeezing $\xi_l$ is a stronger entanglement witness than the global Fisher information $f_G$.}
\label{fig:fullion}
\end{figure}

The complete spectrum of the entanglement coefficients can be revealed if a transverse field is switched on close to the critical point, e.g., $B=J_0$. To account for possible incoherent effects that may occur in a realistic ion-based quantum simulation we additionally consider spin relaxation and dephasing noise induced by spontaneous emission \cite{Bohnet1297,PhysRevA.87.042101}. This is described by means of a Lindblad master equation \cite{BreuerPetruccione2006} with $3N$ decay channels as
\begin{align}\label{eq:noisydyn}
\frac{\partial}{\partial t}\hat{\rho}=-i[\hat{H}_{\alpha,B},\hat{\rho}]+\sum_{l=1}^3\sum_{i=1}^N\gamma_l\left(\hat{L}^{(l)}_i\hat{\rho}\hat{L}^{(l)\dagger}_i-\frac{1}{2}\{\hat{L}^{(l)\dagger}_i\hat{L}^{(l)}_i,\hat{\rho}\}\right)
\end{align}
with Lindblad operators $\hat{L}^{(1)}_i=\hat{\sigma}_i^{(-)}$, $\hat{L}^{(2)}_i=\hat{\sigma}_i^{(+)}$, $\hat{L}^{(3)}_i=\hat{\sigma}_i^{(z)}$ and decay rates $\gamma_1=\gamma_2=\gamma_3/8$ determined by $\gamma = (\gamma_1 + \gamma_2 + \gamma_3)/2$ \cite{PhysRevA.87.042101}. We introduced the ladder operators $\hat{\sigma}_i^{(\pm)} = (\hat{\sigma}_i^{(x)}\pm i\hat{\sigma}_i^{(y)})/2$. 

For the simulation in Fig.~\ref{fig:fullion}, we chose the parameters $B=J_0$, $\alpha=0.2$, and $\gamma=0.01J_0$ and the initial state $|\Psi_0\rangle$, Eq.~(\ref{eq:asyminit}). The depicted complex dynamical evolution can be attributed to the value of $B$ in the vicinity of the quantum phase transition \cite{PhysRevB.93.155153}. It further illustrates the hierarchy among the coefficients which was summarized in Fig.~\ref{fig:hierarchy}. In particular, we note that for a finite time interval the local spin-squeezing coefficient $\xi_l$ is able to detect entanglement of non-Gaussian states, which is not revealed by the global spin-squeezing coefficients. Most remarkably, for short propagation times this coefficient exceeds the Fisher density $f_G$ for global, collective rotations.

\section{Summary and Conclusions}
To summarize, we have developed a unified approach to entanglement detection in multipartite systems. The strongest criteria within our framework are the coefficients~(\ref{eq:genFD}), which are derived from Eq.~(\ref{eq:sepcriterion}) and involve measurements of the Fisher information and local variances \cite{Gessner2016}. Further coefficients based on Eq.~(\ref{eq:variancecriterion}) were given in terms of local squeezing coefficients~(\ref{eq:genSC}) obtained from first and second moments of generic observables. These techniques can be implemented to improve the entanglement detection in a variety of experiments with discrete and continuous variables, by making use of locally resolved access to the system. 

The detailed study of the $N$-qubit case in Sec.~\ref{sec:qubits} highlights the role of local observables for entanglement detection, 
in particular for strongly asymmetric quantum states. 
The local spin-squeezing coefficients can be measured in spin systems with single-site resolution and access to second moments, such as trapped-ion systems \cite{Richerme2014,Jurcevic2014}, or cold atoms under quantum gas microscopes \cite{Parsons2016,Boll2016,Cheuk2016}. As illustrated with examples, these coefficients can lead to a stronger entanglement witness than the globally measured Fisher information, as used in \cite{PhysRevLett.102.100401,Strobel424,Bohnet1297}. Furthermore, they are able to detect the entanglement of states that are no longer recognized by the standard definition of the spin-squeezing coefficient due to their non-Gaussian nature. 

Aside from providing a general way to witness entanglement in
a many-body system, our results and methods have direct implications for quantum metrology using local transformations.
In particular, we have derived the phase sensitivity bound for
separable states when the phase shift is generated by an inhomogeneous spin operator.

The general form of the coefficients~(\ref{eq:genFD}) and~(\ref{eq:genSC}) further permits the development of Fisher densities and squeezing coefficients for continuous-variable systems \cite{BosonicSqueezing} or hybrid systems of discrete and continuous variables \cite{Jeong2014,Morin2014}. 

The coefficients discussed here detect entanglement without specifying which of the subsystems are entangled. A suitable generalization of the separability condition~(\ref{eq:sepcriterion}) may further be used to generalize the coefficients introduced here in order to witness entanglement in a particular partition of the multipartite system \cite{BosonicSqueezing}.

\acknowledgments
The simulations were performed with the help of the \texttt{qutip} package \cite{qutip}. M.G. acknowledges support from the Alexander von Humboldt Foundation.

\appendix
\section{Local spin-squeezing coefficients for inhomogeneous probing}\label{app:lclss}
In this Appendix, we derive two inhomogeneous spin-squeezing coefficients from Eq.~(\ref{eq:localspinsqueezevectors}) and discuss their place in the hierarchy in Fig.~\ref{fig:hierarchy}.

\subsection{Inhomogeneous local spin-squeezing coefficients}
Instead of the normalized mean-spin vector~(\ref{eq:localmeanspin}), the coefficients in this section are based on the non-normalized mean-spin vector
\begin{align}\label{eq:nonormmeanspin}
\mathbf{c}^L_0=(\mathbf{m}_1,\dots,\mathbf{m}_N).
\end{align}
Recall from Sec.~\ref{sec:generallocalqubit} that $\mathbf{c}^L_0$ is not necessarily locally normalized since the length $|\mathbf{m}_i|^2$ 
of the local spin vector represents the purity of the $i$th spin. 

In Eq.~(\ref{eq:localspinsqueezevectors}), we now choose locally orthogonal vectors $\mathbf{c}_1=(\mathbf{n}_1,\dots,\mathbf{n}_N)$ and $\mathbf{c}_2=(\mathbf{n}'_1,\dots,\mathbf{n}'_N)$, such that $\mathbf{c}_3=\mathbf{c}^L_0$. Since Eq.~(\ref{eq:localVarOrthSpinSqueeze}) holds for any vector $\mathbf{c}_1$ that is locally orthogonal to $\mathbf{c}^L_0$, we have $4\mathrm{Var}\big(\hat{A}(\mathbf{c}_1)\big)_{\Pi(\hat{\rho})}=|\mathbf{c}_1|^2$. Furthermore, the condition $\mathbf{m}_i=\mathbf{n}_i\times\mathbf{n}'_i$ [recall Eq.~(\ref{eq:AofCCommutators})] together with the local orthogonality of $\mathbf{c}_1$ and $\mathbf{c}_2$ implies that $|\mathbf{n}_i|^2=|\mathbf{m}_i|^2/|\mathbf{n}'_i|^2$. Hence, $|\mathbf{c}_1|^2=\sum_{i=1}^N|\mathbf{n}_i|^2$ is fully determined by $\mathbf{c}^L_0$ and $\mathbf{c}_2$, leaving only the vector $\mathbf{c}_2$ as a tunable parameter. Thus, we define the inhomogeneous local spin-squeezing coefficient as
\begin{align}\label{eq:xiL}
\xi^2_{L}(\hat{\rho}):=\min_{\tilde{\mathbf{c}}_{\perp}}\frac{\left(\sum_{i=1}^N\big|\frac{\mathbf{m}_i}{\tilde{\mathbf{n}}_i}\big|^2\right)\mathrm{Var}(\hat{A}(\tilde{\mathbf{c}}_{\perp}))_{\hat{\rho}}}{\langle \hat{A}(\mathbf{c}^L_0)\rangle_{\hat{\rho}}^2},
\end{align}
where the minimum is performed over vectors $\tilde{\mathbf{c}}_{\perp}=(\tilde{\mathbf{n}}_1,\dots,\tilde{\mathbf{n}}_N)$ that are locally orthogonal to $\mathbf{c}_0$ and have nonzero local components in all $N$ subsystems, i.e., $|\tilde{\mathbf{n}}_i|^2\neq 0$ for all $i=1,\dots,N$. The spin expectation value along the mean-spin direction can further be expressed in terms of the vector~(\ref{eq:nonormmeanspin}) as
\begin{align}\label{eq:nonormmeanspinexpval}
\langle\hat{A}(\mathbf{c}^L_0)\rangle_{\hat{\rho}}=\frac{1}{2}\sum_{i=1}^N\mathbf{m}_i\cdot\langle\hat{\boldsymbol{\sigma}}_i\rangle_{\hat{\rho}}=\frac{1}{2}\sum_{i=1}^N|\mathbf{m}_i|^2=\frac{|\mathbf{c}^L_0|^2}{2}.
\end{align}

As a special case of Eq.~(\ref{eq:xiL}), we minimize only over locally normalized vectors $\mathbf{c}_{\perp}$. This leads to the partially inhomogeneous local spin-squeezing coefficient:
\begin{align}\label{eq:xiLl}
\xi^2_{Ll}(\hat{\rho}):=\min_{\mathbf{c}_{\perp}}\frac{2\mathrm{Var}(\hat{A}(\mathbf{c}_{\perp}))_{\hat{\rho}}}{\langle \hat{A}(\mathbf{c}^L_0)\rangle_{\hat{\rho}}}.
\end{align}
Here, we used Eq.~(\ref{eq:nonormmeanspinexpval}), which together with Eq.~(\ref{eq:localVarOrthSpinSqueeze}) and $|\mathbf{c}_{1}|^2=\sum_{i=1}^N|\mathbf{m}_i|^2$ implies that $2\mathrm{Var}\big(\hat{A}(\mathbf{c}_{1})\big)_{\Pi(\hat{\rho})}=\langle\hat{A}(\mathbf{c}^L_0)\rangle_{\hat{\rho}}$. 

We used the subscript $Ll$ to indicate that Eq.~(\ref{eq:xiLl}) contains both a locally non-normalized vector (the mean spin $\mathbf{c}_0^L$; subscript $L$) and a locally normalized vector (optimization over $\mathbf{c}_{\perp}$; subscript $l$). In Eq.~(\ref{eq:xiL}) the optimization is also performed over a locally non-normalized vector, hence the subscript $L$.

The optimization involved in $\xi_L$ incorporates $2N$ parameters, whereas the additional normalization constraints reduce this number to $N$ for $\xi_{Ll}$.

\subsection{Relation to other coefficients}
Let us now discuss the relations among the local spin-squeezing coefficients $\xi_L$, $\xi_{Ll}$, and $\xi_l$ defined in Eqs.~(\ref{eq:xiL}), (\ref{eq:xiLl}), and (\ref{eq:xil}). First note that all the spin-squeezing coefficients in this paper are special cases of Eq.~(\ref{eq:localvariancecoefficient}) and~(\ref{eq:localspinsqueezevectors}) and therefore generate a separability criterion~(\ref{eq:squeezingbound}). The two coefficients presented in this Appendix are both stronger entanglement criteria than the local spin-squeezing coefficient $\xi_l$ that was discussed in the main text. We have
\begin{align}
\xi_L^{-2} \geq \xi_{Ll}^{-2} \geq \xi_l^{-2}.
\end{align}

The first inequality follows directly since $\xi_{Ll}$ is obtained by imposing additional local normalization constraints on $\xi_L$. For the second relation, note that we can use Eqs.~(\ref{eq:nonormmeanspinexpval}) and~(\ref{eq:normmeanspinexpval}) to rewrite the coefficients as
\begin{align}
\xi_{Ll}^{2}(\hat{\rho})&= \frac{4}{\sum_{i=1}^N|\mathbf{m}_i|^2}\min_{\mathbf{c}_{\perp}}\mathrm{Var}(\hat{A}(\mathbf{c}_{\perp}))_{\hat{\rho}}
\end{align}
and
\begin{align}
\xi_l^{2}(\hat{\rho})=\frac{4N}{\left(\sum_{i=1}^N|\mathbf{m}_i|\right)^2}\min_{\mathbf{c}_{\perp}}\mathrm{Var}(\hat{A}(\mathbf{c}_{\perp}))_{\hat{\rho}}.
\end{align}
The relation $\xi_{Ll}^{2}\leq\xi_l^{2}$ now follows by virtue of the Cauchy-Schwarz inequality $N\sum_{i=1}^N|\mathbf{m}_i|^2\geq (\sum_{i=1}^N|\mathbf{m}_i|)^2$.

Under certain conditions, we can further bound the Fisher density $f_L$ by the spin-squeezing coefficient $\xi_L$ as
\begin{align}
f_L(\hat{\rho}) \geq \xi_L^{-2}(\hat{\rho}).
\end{align}
To see this, we write
\begin{align}
f_L(\hat{\rho})&=\frac{F_Q[\hat{\rho},\hat{A}(\mathbf{c}^{L}_{\mathrm{opt}})]}{|\mathbf{c}^L_{\mathrm{opt}}|^2}\notag\\
 &\geq \max_{\mathbf{c}'}\frac{|\langle [\hat{A}(\mathbf{c}^L_{\mathrm{opt}}),\hat{A}(\mathbf{c}')]\rangle_{\hat{\rho}}|^2}{|\mathbf{c}^L_{\mathrm{opt}}|^2\mathrm{Var}(\hat{A}(\mathbf{c}'))_{\hat{\rho}}}.
\end{align}
We can now restrict the maximization to these vectors $\mathbf{c}=(\mathbf{n}_1,\dots,\mathbf{n}_N)$ that are locally orthogonal to $\mathbf{c}^L_{\mathrm{opt}}$ and satisfy the conditions $\mathbf{m}_i=\mathbf{n}^L_{\mathrm{opt,i}}\times\mathbf{n}_{i}$ for all $i=1,\dots,N$. This requires that $|\mathbf{n}^L_{\mathrm{opt,i}}|=|\mathbf{m}_i|/|\mathbf{n}_{i}|$, and with $|\mathbf{c}^L_{\mathrm{opt}}|^2=\sum_{i=1}^N|\mathbf{n}^L_{\mathrm{opt,i}}|^2$ we obtain the bound
\begin{align}
f_L(\hat{\rho})&\geq \max_{\mathbf{c}}\frac{\langle \hat{A}(\mathbf{c}^L_0)\rangle_{\hat{\rho}}^2}{\left(\sum_{i=1}^N\big|\frac{\mathbf{m}_i}{\mathbf{n}_i}\big|\right)^2\mathrm{Var}(\hat{A}(\mathbf{c}))_{\hat{\rho}}}\notag\\
&=\xi_L^{-2}(\hat{\rho}).
\end{align}
This bound, however, holds only for states with the property $|\mathbf{m}_i|^2=0\:\Leftrightarrow\:|\mathbf{n}^L_{\mathrm{opt,i}}|^2=0$.

Following the arguments in Secs.~\ref{sec:metro} and~\ref{sec:multi}, we conclude that both $\xi_{L}$ and $\xi_{Ll}$ can be interpreted in terms of a metrological quantum gain in the case of inhomogeneous probing, but in contrast to $\xi_l$, it is presently unknown if they can also quantify multipartite entanglement.


\begin{thebibliography}{64}
\bibitem [{foo({\natexlab{a}})}]{footnote1}%
  \BibitemOpen
  \href@noop {} {}\bibinfo {note} {To ensure that
  $f^V_{\mathbf{c}}(\hat{\rho})<\infty$, we must pay special attention to the
  case of a pure product state, which, as discussed in
  Sec.~\ref{sec:generallocalqubit} is the only state for which the denominator
  may potentially vanish. Such states satisfy the exact equality
  $F_Q[|\Psi_1\rangle\otimes\cdots\otimes|\Psi_N\rangle,\hat{A}(\mathbf{c})]=4\mathrm{Var}(\hat{A}(\mathbf{c}))_{|\Psi_1\rangle\otimes\cdots\otimes|\Psi_N\rangle}$
  for all $\mathbf{c}$. Hence, in this case we assign the value
  $f^V_{\mathbf{c}}(|\Psi_1\rangle\otimes\cdots\otimes|\Psi_N\rangle)=1$.}\BibitemShut
  {Stop}%
\bibitem [{foo({\natexlab{b}})}]{footnote2}%
  \BibitemOpen
  \href@noop {} {}\bibinfo {note} {A rather obvious choice
  for the optimization of $f^V_{\mathbf{c}}$ would be
  $\tilde{f}^V_L(\hat{\rho}):=\max_{\mathbf{c}}f^V_{\mathbf{c}}(\hat{\rho})$.
  Yet $\tilde{f}^V_L$ is more difficult to compute than $\tilde{f}^V_L$ while
  leading to an equally powerful entanglement criterion. Indeed, we have
  $\tilde{f}^V_L(\hat{\rho})>1\quad\Leftrightarrow\quad f^V_L(\hat{\rho})>1$.
  To see this, first notice that $\Leftarrow$ follows immediately since
  $\tilde{f}^V_L(\hat{\rho})\geq f^V_L(\hat{\rho})$ is always true. The
  $\Rightarrow$ follows from the fact that $\tilde{f}^V_L(\hat{\rho})>1$
  implies that there exists at least one $\mathbf{c}$ such that
  $F_Q[\hat{\rho},\hat{A}(\mathbf{c})]>4\mathrm{Var}(\hat{A}(\mathbf{c}))_{\Pi(\hat{\rho})}$,
  and consequently, also $f^V_L(\hat{\rho})>1$ must be true. Hence, there is no
  additional gain by performing the more complicated optimization for
  $\tilde{f}^V_L$, and henceforth, we limit our analysis to the coefficient
  $f^V_L$.}\BibitemShut {Stop}%
\end{thebibliography}
\end{document}